\documentclass[aps,prd,amsmath,amssymb,10pt,letterpaper,balancelastpage,showpacs,superscriptaddress,nofootinbib,floatfix,notitlepage,longbibliography,twocolumn]{revtex4-2}

\usepackage{url}
\usepackage{color}
\usepackage[usenames,dvipsnames,svgnames,table]{xcolor}
\usepackage{mathrsfs}

\usepackage{graphicx}	
\usepackage{ragged2e}	
\usepackage{bm}		    
\usepackage[sort&compress]{natbib}	 	
	\setcitestyle{square,numbers,comma}	
\usepackage[colorlinks=true,urlcolor=blue,linkcolor=black,citecolor=red]{hyperref}	

\newcommand{\orcid}[1]{\href{https://orcid.org/#1}{\,\includegraphics[width=8px]{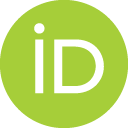}}}

\newcommand{\tabref}[2][]{Tab{#1}.~\ref{#2}}		
\newcommand{\figref}[2][]{Fig{#1}.~\ref{#2}}		
\newcommand{\sectref}[2][]{Sec{#1}.~\ref{#2}}		
\newcommand{\appref}[2][x]{Appendi{#1}~\ref{#2}}	
\renewcommand{\eqref}[2][]{Eq{#1}.~(\ref{#2})}		
\newcommand{\eqrefRange}[2]{Eqs.~(\ref{#1})--(\ref{#2})}		
\newcommand{\citeR}[2][]{Ref{#1}.~\cite{#2}}			
\newcommand{\lb}{\ensuremath{\left}}					
\newcommand{\rb}{\ensuremath{\right}}					
\newcommand{\order}[1]{\ensuremath{\mathcal{O}(#1)}}    
\newcommand{\Eros}{433~Eros}
\newcommand{\Thomsen}{2064~Thomsen}
\newcommand{\cl}{\text{cl}}
\newcommand{\mcl}{\ensuremath{m_{\cl}}}
\newcommand{\vcl}{\ensuremath{v_{\cl}}}

\newcommand{\hcl}{\ensuremath{h_{\cl}}}
\newcommand{\dm}{\textsc{dm}}
\newcommand{\gw}{\textsc{GW}}

\begin{document}

\title{Searching for Dark Clumps with Gravitational-Wave Detectors}
\date{\today}

\author{Sebastian Baum\orcid{0000-0001-6792-9381}}
\email{sbaum@stanford.edu}
\affiliation{Stanford Institute for Theoretical Physics, Department of Physics, Stanford University, Stanford, CA 94305, USA}
\author{Michael A.~Fedderke\orcid{0000-0002-1319-1622}}
\email{mfedderke@jhu.edu}
\affiliation{The William H.~Miller III Department of Physics and Astronomy, The Johns Hopkins University, Baltimore, MD  21218, USA}
\author{Peter W.~Graham\orcid{0000-0002-1600-1601}\,}
\email{pwgraham@stanford.edu}
\affiliation{Stanford Institute for Theoretical Physics, Department of Physics, Stanford University, Stanford, CA 94305, USA}
\affiliation{Kavli Institute for Particle Astrophysics \& Cosmology, Department of Physics, Stanford University, Stanford, CA 94305, USA}

\begin{abstract}
Dark compact objects (``clumps'') transiting the Solar System exert accelerations on the test masses (TM) in a gravitational-wave (GW) detector.
We reexamine the detectability of these clump transits in a variety of current and future GW detectors, operating over a broad range of frequencies.
TM accelerations induced by clump transits through the inner Solar System have frequency content around $f \sim \mu$Hz. 
Some of us [Fedderke \emph{et al}., \href{https://doi.org/10.1103/PhysRevD.105.103018}{Phys.~Rev.~D \textbf{105}, 103018 (2022)}] recently proposed a GW detection concept with $\mu$Hz sensitivity, based on asteroid-to-asteroid ranging.
From the detailed sensitivity projection for this concept, we find both analytically and in simulation that purely gravitational clump--matter interactions would yield one detectable transit every $\sim20$\,yrs, if clumps with mass $\mcl \sim 10^{14}\,\text{kg}$ saturate the dark-matter (DM) density.
Other~(proposed) GW detectors using local TMs and operating in higher frequency bands are sensitive to smaller clump masses and have smaller rates of discoverable signals. 
We also consider the case of clumps endowed with an additional attractive long-range clump--matter fifth force significantly stronger than gravity~(but evading known fifth-force constraints). 
For the $\mu$Hz detector concept, we use simulations to show that, for example, a clump--matter fifth-force $\sim 10^3$ times stronger than gravity with a range of~$\sim\text{AU}$ would boost the rate of detectable transits to a few per year for clumps in the mass range~$10^{11}\,\text{kg} \lesssim \mcl \lesssim 10^{14}\,\text{kg}$, even if they are a $\sim 1\,$\% sub-component of the DM.
The ability of $\mu$Hz GW detectors to probe asteroid-mass-scale dark objects that may otherwise be undetectable bolsters the science case for their development.
\end{abstract}

\maketitle

\tableofcontents
\pagebreak

\section{Introduction}
\label{sect:introduction}

The direct detection of gravitational waves (GWs) by the LIGO/Virgo Collaborations~\cite{LIGOScientific:2016aoc,PhysRevLett.119.161101} has opened a new window through which to observe the Universe. 
The \order{100} compact-object merger events that have since been observed~\cite{LIGOScientific:2018mvr,LIGOScientific:2021usb,LIGOScientific:2021djp} have provided insight into such varied topics as massive black-hole populations and formation histories~\cite{LIGOScientific:2020kqk}; the synthesis of heavy elements in the Universe~\cite{LIGOScientific:2017ync}; precision measurements of the speed of propagation of gravitational disturbances~\cite{LIGOScientific:2017zic} and of dynamical, strong-field general relativity~\cite{LIGOScientific:2018dkp,LIGOScientific:2021sio}; and new ways to measure the Hubble constant~\cite{LIGOScientific:2017adf,Chen:2017rfc,LIGOScientific:2018gmd,Doctor:2019odr,DES:2019ccw}.
While ground-based laser interferometers such as the LIGO/Virgo/KAGRA~\cite{KAGRA:2020cvd} network are sensitive to GWs in the $\sim 10\,$Hz--10\,kHz frequency band, existing pulsar timing arrays (PTAs)~\cite{Kramer:2013kea,Shannon:2015ect,Verbiest:2016vem,Kerr:2020qdo,NANOGrav:2020bcs} are sensitive to nHz--$\mu$Hz GW signals. 
These PTAs are also beginning to observe an interesting common-spectrum process~\cite{NANOGrav:2020bcs,Goncharov:2021oub,2021MNRAS.508.4970C,2022MNRAS.510.4873A}, which could be the herald of further discoveries to come.

A wide array of concepts is proposed and/or under development to cover as much as possible of the GW frequency range between PTAs and ground-based laser interferometers, including ideas based on astrometry~\cite{Pyne:1995iy,Schutz:2010lmv,Book:2010pf,Klioner:2017asb,Moore:2017ity,Park:2019ies,Wang:2020pmf,Fedderke:2022kxq,Wang:2022sxn}, ranging between asteroids~\cite{Fedderke:2021kuy}, studying orbital perturbations of astrophysical binaries~\cite{Blas:2021mqw,Blas:2021mpc}, and future atomic~\cite{Dimopoulos:2007cj,Hogan:2011tsw,Canuel:2017rrp,Graham:2017pmn,Zhan:2019quq,Tino:2019tkb,AEDGE:2019nxb,Badurina:2019hst,MAGIS-100:2021etm} or laser~\cite{TianQin:2015yph,TianQin:2020hid,LISA:2017pwj,Armano:2016bkm,Armano:2016bkm,Baker:2019nia,Maggiore:2019uih,Reitze:2019iox,Sesana:2019vho,Kawamura:2020pcg} interferometers on Earth or in space. 
Other space-based proposals include using atomic clocks on orbit~\cite{Kolkowitz:2016wyg,Alonso:2022oot} and deploying lunar GW detectors of various types~\cite{LSGreport,PAIK2009167,PhysRevD.90.102001,LGW-ESA,LGWA:2020mma,LSGA,Jani:2020gnz}; studies of ground-based detectors with coverage at higher (MHz--GHz) frequencies have also been undertaken, see, e.g., \citeR{Aggarwal:2020olq} for a recent review.
While many exciting lessons await from further observations of merging binaries (e.g., information regarding gaps and structure in the compact-object mass distribution~\cite{LIGOScientific:2021psn,Farah:2021qom,Ye:2022qoe,Edelman:2021fik}, including implications for new physics~\cite{Croon:2020oga,Baxter:2021swn}) or future measurements of stochastic GW signals, it also behooves us to consider other discovery opportunities that existing and proposed GW detectors enable.

In this work, we revisit an old idea~\cite{Seto:2004zu,Adams:2004pk} to use GW detectors to search for the nearby passage of a massive, compact object (see also \citeR[s]{Ni:2002cd,Vinet:2006fj,Thorpe:2015cxa,LISA-Pre-Phase-A,Hall:2016usm}). 
Such a passage would induce a time-dependent change to the test-mass (TM) separation in the GW detector via the gravitational (or any additional force) interactions between that object and the TMs.
For concreteness, we will consider the sensitivity of GW detectors to macroscopic compact dark matter (DM) ``clumps'' in the broad mass range of $1\textup{--}10^{20}\,{\rm kg} \sim  10^{-30}\textup{--}10^{-10}\,M_\odot$; see also \citeR[s]{Seto:2007kj,Siegel:2007fz,Schutz:2016khr,Dror:2019twh,Ramani:2020hdo} for related work on searching for DM clumps at the upper end of this mass range with PTAs, \citeR{Kawasaki:2018xak} for searches in the kg-range with ground-based laser interferometers, and \citeR{Das:2021drz} for a proposal to search for $10^{13}\textup{--}10^{19}\,{\rm kg} \sim 10^{-17}\textup{--}10^{-11}\,M_\odot$ DM clumps via stellar shocks. 
Our analysis is insensitive to the nature of these clumps, and encompasses possibilities such as (primordial) black holes~\cite{Khlopov:2008qy,Carr:2009jm,Lehoucq:2009ge,Clark:2016nst,Carr:2016drx,Lehmann:2018ejc,Katz:2018zrn,DeRocco:2019fjq,Smyth:2019whb,Lehmann:2019zgt,Kashlinsky:2019kac,Montero-Camacho:2019jte,Carr:2020xqk,Carr:2020gox,Laha:2020ivk,Lehmann:2020bby,Franciolini:2022htd,Chakraborty:2022mwu} and composite DM objects like blobs/nuggets~\cite{Wise:2014jva,Wise:2014ola,Hardy:2014mqa,Hardy:2015boa,Gresham:2017zqi,Gresham:2017cvl,Gresham:2018anj,Bai:2018dxf,Coskuner:2018are,Grabowska:2018lnd,Hong:2020est,Diamond:2021dth}, as long as they are stable and dense enough not to be tidally disrupted during their passage through the inner Solar System. 
Our analysis also readily extends to other, non-DM objects: e.g., rogue asteroids such as `Oumuamua.
Our work is similar in conception to that carried out in \citeR{Hall:2016usm}, which considered LISA and aLIGO sensitivity to the passage of dark clumps (both for purely gravitational and fifth-force couplings); our main innovation is in considering the sensitivity of more recently proposed detection concepts that would expand the mass range over which such clumps could be searched for in this way (we also revisit the LISA sensitivity using current LISA projections).

DM clumps and other interstellar objects have typical speeds of order $\vcl \sim 10^{-3}\,c$ relative to the Solar System.
The transit of such objects through the inner few-AU%
\footnote{\label{ftnt:AUdefn}%
    $1\,\text{AU} \approx 1.5\times 10^{11}\,\text{m}$.
    } %
region of the Solar System takes $T \sim 10^6\,$s, placing the frequency content of the corresponding signal in a GW detector squarely in the $0.1\textup{--}10\,\mu$Hz band.
Refs.~\cite{Seto:2004zu,Adams:2004pk} considered a LISA-like detector, but used overly optimistic estimates for the low-frequency noise. In particular, the stability of human-engineered TMs in space (at least those feasible with state-of-the-art technology) severely limits the sensitivity of such a detector below the mHz range, as demonstrated by the LISA Pathfinder mission~\cite{Armano:2016bkm,Armano:2018kix}. 
Much smaller low-frequency acceleration noise of local TMs could potentially be achieved for TMs that are much more massive than those that can be launched from the ground. 
\citeR{Fedderke:2021kuy} showed that certain asteroids in the inner Solar System can be sufficiently stable to serve as excellent TMs for GW searches in the $\mu$Hz band, with the GW detection implemented by direct asteroid-to-asteroid ranging via deployed laser or radio links. 
For carefully chosen asteroids, the dominant noise source limiting the stability of the asteroids as TMs at frequencies $\lesssim \mu$Hz is the so-called ``asteroid gravity gradient noise'', which was characterized in \citeR{Fedderke:2020yfy}. 
This fundamental, unshieldable noise source~(the computation of which in \citeR{Fedderke:2020yfy} applies to all GW detectors with both ends of the baseline confined to the inner Solar System) 
is due to the gravitational acceleration of the TMs induced by the~$\mathcal{O}(10^6)$ other inner Solar System asteroids.
Its origin is hence conceptually similar to the TM acceleration signal arising from the close passage of a compact object we are attempting to search for. 
The recently improved understanding of the relevant noise sources in the $\mu$Hz frequency band~\cite{Fedderke:2020yfy} as well as the rich diversity of existing and proposed local-TM-based GW detector concepts across the GW frequency landscape from $\mu$Hz up to kHz and beyond makes this an opportune moment to reconsider the sensitivity of GW detectors to DM clumps.\\

In \sectref{sect:OoM} we discuss an heuristic estimate of the sensitivity, through purely gravitational interactions, of GW detectors to the passage of compact objects through the inner Solar System. 
This estimate allows us to compare the reach of different detectors and to understand why a $\mu$Hz GW detector, such as one based on asteroid-to-asteroid ranging~\cite{Fedderke:2021kuy}, is particularly well-suited for such a search.
Assuming that the clumps make up a fixed fraction (or all) of the DM, we show that the higher the frequency at which a detector operates, the smaller the clump masses it is sensitive to. 
Due to the parametric scaling of the strain sensitivity of proposed detectors with GW frequency and baseline length, among the detectors and concepts that we consider, the rate of discoverable signals is highest for the asteroid-ranging proposal. 
A detector operating in the $\mu$Hz range would be most sensitive to clumps in the $10^{13}\textup{--}10^{16}\,$kg range. 
Taking the estimated sensitivity from Ref.~\cite{Fedderke:2021kuy} for the asteroid-ranging concept, the rate of discoverable signals could be as large as  $\sim 5\times 10^{-2}\,{\rm yr}^{-1}$ for clump masses $\sim 10^{14}\,$kg, assuming that these clumps make up all of the DM. 
Detectors operating in the kHz band (e.g., ground-based laser interferometers) are best suited to search for clumps with masses in the $10\textup{--}10^4\,$kg mass range; the rate of detectable signals in a detector with the sensitivity of the Einstein Telescope~\cite{Maggiore:2019uih} or Cosmic Explorer~\cite{Reitze:2019iox} proposals could be as large as $2 \times 10^{-2}\,{\rm yr}^{-1}$ for $\sim 10^2\,$kg clumps making up the DM. Detectors operating at intermediate frequencies, such as space-based laser~(e.g., LISA~\cite{LISA:2017pwj} or TianQin~\cite{TianQin:2015yph}) or atom interferometers, would be most sensitive to clump masses in between these ranges, but the expected rate of discoverable signals tends to be smaller than those for the asteroid ranging or next-generation ground-based laser-interferometer proposals. However, detectors operating in different frequency ranges are complimentary, as they are sensitive to clumps of different masses.

In \sectref{sect:MF} we then focus on the asteroid-ranging concept of \citeR{Fedderke:2021kuy} and present an improved sensitivity estimate based on a matched-filter search, again assuming purely gravitational interactions of the transiting compact object and the detector TMs. 
We present the results of a Monte Carlo simulation we implemented in order to calculate the sensitivity.
This simulation accounts for the distribution of trajectories that DM clumps would take through the inner Solar System.
We consider both a toy example of ranging two (fictitious) asteroids located $0.5\,$AU from the Sun and separated from each other by 1\,AU, and a more realistic setup of ranging between two known asteroids with appropriate properties to serve as good TMs, \Eros\ and \Thomsen.
The projected sensitivities we find from this analysis are in good agreement with our heuristic estimates in \sectref{sect:OoM}.

In \sectref{sect:FF} we entertain the possibility of a stronger-than-gravity long-range fifth force between SM particles and DM clumps. 
Such a fifth force would have multiple interesting effects: while the force could obviously enhance the signal from a passing clump in the asteroid GW detector, there would also be a relevant focusing effect on the clump trajectories through the inner Solar System due to the fifth-force interaction between the clumps and the Sun. 
The latter effect reinforces the former, leading overall to a vastly enhanced detectable rate of transits, as high as a few per year for some parameter choices, even assuming that the clumps comprise only a \order{1\%} fraction of the DM.

We conclude in Sec.~\ref{sect:conclusions}.

\section{Heuristic Sensitivity Estimate}
\label{sect:OoM}

Let us discuss a rough estimate of the sensitivity of a GW detector to the passage of a clump in order to gain some intuitive understanding of the sensitivity. 
As noted above, this search will be insensitive to the nature of these clumps as long as they are stable and dense enough not to be tidally disrupted during their passage through the inner Solar System. 
Throughout this work, we denote the mass of the clump by $\mcl$, and its speed relative to the detector by $\vcl$.
The acceleration%
\footnote{\label{ftnt:Shapiro}%
    It is well known that the Shapiro effect resulting from a clump passage anywhere along the baseline also gives a potential signal~\cite{Siegel:2007fz}. We have checked that, for the asteroid-ranging proposal of \citeR{Fedderke:2021kuy} that we study most carefully in this work, the rate of detectable DM clump transits via the Shapiro effect is highly suppressed as compared to the direct acceleration effect. 
    This is because the clumps must be much more massive to generate a detectable Shapiro-delay signal.
    With very long baselines, such as in PTAs, close passages anywhere along the baseline can compensate the rate loss from the more massive clump (see, e.g., discussion in \citeR{Dror:2019twh}); for the baselines we consider here, however, such a compensation does not occur.
    } %
such a clump will exert on a TM when the distance of the clump to the TM is $r$ is then
\begin{align} \label{eq:acc}
    a \sim \frac{G_N \mcl}{r^2} \:,
\end{align}
where $G_N$ is the gravitational constant. 
A GW detector does not measure the absolute acceleration of a TM, but rather the relative acceleration between two TMs. 
If we denote the separation of the TMs by $L$, the relative acceleration between the TMs from the passage of a clump will be
\begin{align} \label{eq:Da}
    \Delta a \sim G_N \mcl \left(\frac{1}{r_1^2} - \frac{1}{r_2^2}\right) \sim \frac{G_N \mcl}{r^2} \min\left( 1, \frac{2 L}{r} \right) \:,
\end{align}
where $r_i$ is the distance of the clump to the $i$-th TM. 
In the second expression in \eqref{eq:Da}, $r$ denotes the distance between the clump and the closest TM.
The precise form of the appropriate approximation depends on the relative orientation of the clump trajectory to the TMs; qualitatively, however, the relative acceleration is simply given by the acceleration of the closer TM for $r \ll L$, while for $r \gtrsim L$, one picks up a tidal suppression factor of $\sim L/r$. 
As we will see below, GW detectors will mostly be sensitive to encounters with $r \ll L$.

In order to gain intuition for the sensitivity of GW detectors to such accelerations, let us convert this into a ``strain'' signal; i.e., into the integrated fractional change of the separation between the two TMs.
We note that, when searching for clumps, the acceleration is a more natural quantity to consider than the strain; when discussing the matched-filter search in \sectref{sect:MF}, we will do so in terms of the acceleration.
However, for pedagogical purposes, it useful to think about the signal in terms of the strain. 

\begin{figure*}
    \includegraphics[width=\textwidth]{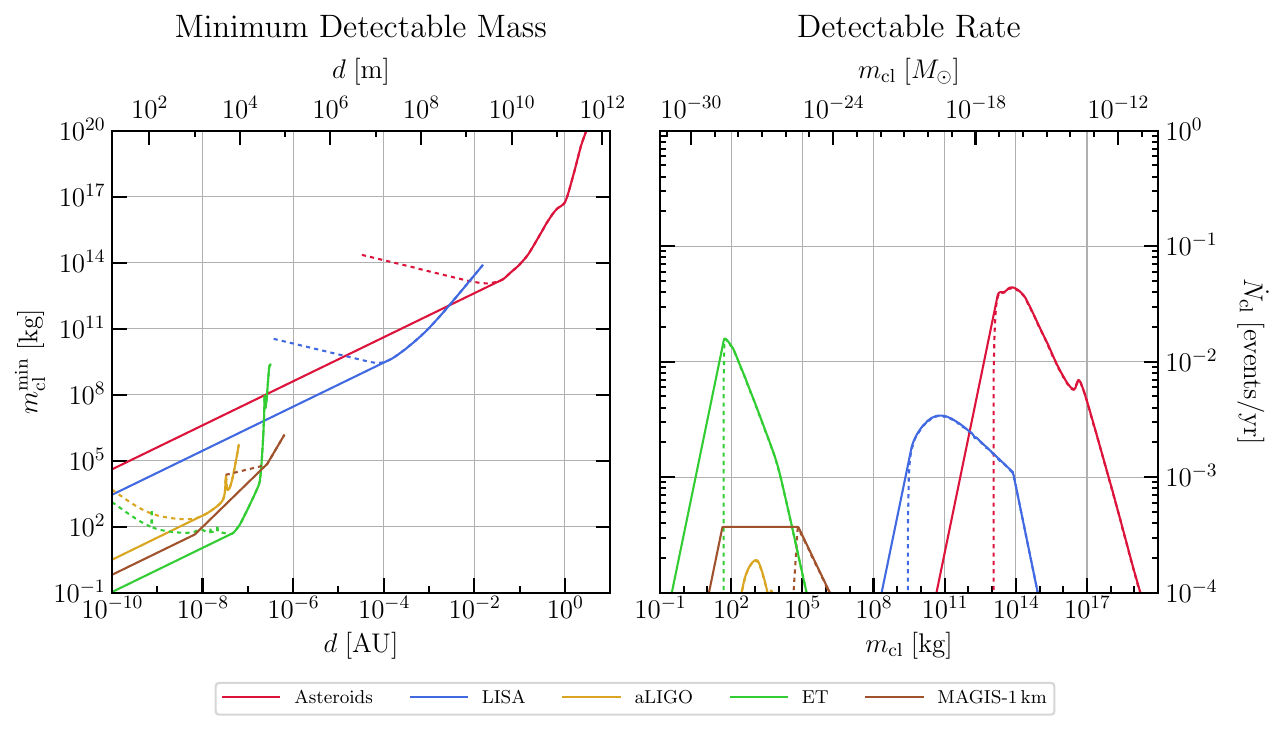}
    \caption{Sensitivity results using the estimate discussed in Sec.~\ref{sect:OoM}. 
    The dashed lines give the estimate based on the strain signal induced at $f_{\rm peak}$ only, while the solid curves show our improved projections taking into account the induced signal at frequencies $f < f_{\rm peak}$, as discussed in the text. 
    The different colors correspond to different existing or proposed GW detectors as indicated in the legend: the asteroid-to-asteroid ranging proposal (``Asteroids''), space-based (LISA) and ground-based [advanced LIGO (aLIGO) and Einstein Telescope (ET)] laser-interferometers, and the ground-based atom-interferometer proposal MAGIS-1km. 
    \textsc{Left panel:}~Smallest clump mass $\mcl^{\rm min}$ for a given point of closest approach to the detectors TMs $d$ that would lead to a discoverable signal assuming a relative speed of the clump to the detector of $\vcl = 300\,$km/s and purely gravitational interactions between the clump and the detector TMs. 
    \textsc{Right panel:}~Rate of discoverable (SNR $\geq 1$) signals $\dot{N}_{\cl}$ assuming that clumps of mass $\mcl$ make up all of the DM, $\rho_{\cl} = 0.3\,$GeV/cm$^3$.}
    \label{fig:OoM_results}
\end{figure*}

Suppose that the clump passes the detector such that the smallest distance between the closest TM and the clump on its trajectory is $d$. 
Given that the clump has speed $\vcl$ relative to the detector, the strain signal will be peaked at the frequency
\begin{align} \label{eq:fpeak}
    f_{\rm peak} \sim \frac{\vcl}{2\pi d} \,.
\end{align}
The strain induced by such a signal is then%
\footnote{\label{ftnt:correctionToHc}%
    Note that care must be taken when comparing this to a GW detector strain sensitivity curve at frequencies $f \gtrsim 1/(\pi L)$~\cite{Morisaki:2020gui}. 
    A GW detector suffers a strain-response suppression in this regime as the GW wavelength falls inside the detector baseline.
    This suppression however does not affect our signal, and so must be removed from the published GW detector sensitivity curves $h_c(f)$ prior to employing them in this work. 
    We have accounted for this effect in drawing the dotted lines in \figref{fig:OoM_results}, where it is relevant. 
    The impact on our other results is however expected to be negligible because (except for edge cases) they are dominated by a region of the sensitivity curves which this correction does not affect; it is thus ignored, which is also a conservative assumption with regard to detectability.
} %
\begin{align} \label{eq:hc}
    \hcl \sim \frac{\Delta a}{ (2\pi f_{\rm peak})^2 L} \sim \frac{G_N \mcl}{\vcl^2 L} \min\left( 1, \frac{4 \pi L f_{\rm peak}}{\vcl} \right) \:. 
\end{align}
Note that as long as $d \ll L$, such that the acceleration is not in the tidal limit (i.e., the ``min'' expression is equal to 1), the induced strain is independent of $d$; this is because the dependence of the relative acceleration of the TMs on $d$ ($\Delta a \propto d^{-2}$) is canceled by the time it takes the clump to pass by the detector ($1/f_{\rm peak} \propto d$). 

We can invert these expressions to derive an heuristic estimate for the sensitivity of a given GW detector. 
A GW detector with baseline $L$ and characteristic strain sensitivity curve $h_c(f)$ is sensitive to clumps coming within a distance
\begin{align} \label{eq:d}
    d \sim \frac{\vcl}{2\pi f} \:,
\end{align}
provided that they have a mass
\begin{align} \label{eq:mcmin}
    \mcl \gtrsim \mcl^{\rm min} = \frac{\vcl^2 h_c(f) L}{G_N} \max\left( 1, \frac{\vcl}{4 \pi L f} \right) \,.
\end{align}
We can use \eqref{eq:d} to eliminate $f$ in favor of $d$ in \eqref{eq:mcmin} and compute $\mcl^{\rm min}(d)$ for any detector.
Assuming the clumps have a local mass density $\rho_{\cl}$, we can estimate the rate of discoverable signals as
\begin{align} \label{eq:Rate}
    \dot{N}_{\cl} \sim \pi d^2 \vcl \frac{\rho_{\cl}}{\mcl^{\rm min}} \sim \frac{G_N \vcl \rho_{\cl}}{4 \pi} \frac{\min\left( 1, \frac{4 \pi L f}{\vcl} \right)}{L f^2 h_c(f)} \:. 
\end{align}
The latter form of this expression is useful to directly calculate the rate of discoverable signals from the detector characteristics, although it obscures that detectors are sensitive to different $\mcl^{\rm min}(d)$, depending on the frequencies at which they operate.

We are now in a position to estimate the sensitivity of different detectors. 
Assuming that the clumps make up the DM, we will use a typical speed of $\vcl = 300\,$km/s and assume a local mass density of $\rho_{\cl} = 0.3\,$GeV/cm$^3$.
To start, let us consider the asteroid-to-asteroid ranging proposal of \citeR{Fedderke:2021kuy}, with a baseline of $L \sim 1\,$AU and optimum strain sensitivity of $h^{\text{opt}}_c \sim 5 \times 10^{-20}$ at $f_{\text{opt}} \sim 10\,\mu$Hz. 
This corresponds to the peak frequency of the signal from a clump coming within $d \sim 0.03\,$AU of the detector (note that $d \ll L$; the TM acceleration signal is thus clearly not in the tidal limit). 
This strain sensitivity corresponds to a minimal detectable clump mass $\mcl^{\rm min} \sim 5 \times 10^{-18}\,M_\odot \approx 10^{13}\,$kg, and we would expect such encounters to occur with a rate of $\dot{N}_{\cl} \sim 0.04\,{\rm yr}^{-1}$. 
However, inspecting \eqref{eq:Rate} we can see that this is not quite the frequency (and, in turn, clump mass) for which we find the largest rate:
since $\dot{N}_{\cl} \propto f^{-2} h_c^{-1}(f)$, the rate continues to increase for $f_{\rm peak} < f_{\text{opt}}$, until a corner frequency $f_{\text{cnr}}$ at which the detector sensitivity curve $h_c(f)$ starts degrading faster than $h_c \propto f^{-2}$.
For the asteroid-ranging proposal, $f_{\text{cnr}} \sim 3\,\mu$Hz, giving $h^{\text{cnr}}_c \sim 5 \times 10^{-19}$. 
Taking $f = f_{\text{cnr}}$ in \eqrefRange{eq:d}{eq:Rate} then gives $\mcl^{\rm min} \sim 5 \times 10^{-17}\,M_\odot \approx 10^{14}\,$kg, $d_{\text{cnr}} \sim 0.1\,$AU, and a signal rate of $\dot{N}_{\cl} \sim 0.05\,{\rm yr}^{-1}$ (i.e., $\sim 1$ detectable event every $\sim 20$\,yrs). It is amusing to note that the asteroid-to-asteroid ranging proposal would be most sensitive to DM clumps with asteroid-like masses.

Let us now compare these results for the asteroid-ranging proposal with other GW detectors. 
All other existing or proposed detectors with all of their TMs located in the Solar System are aimed at higher frequencies.
From \eqref{eq:d}, we can immediately see that a detector optimized for higher-frequency signals will see clumps passing closer to the detector: $d \propto 1/f_{\text{opt}}$.
The optimal characteristic-strain sensitivities of different existing or proposed detectors lie very roughly along a line $h^{\text{opt}}_c \propto f_{\text{opt}}^{-0.7}$.
At the same time, the larger the frequency for which a detector is optimized, the shorter its baseline: approximately, detectors are designed%
\footnote{Atom interferometer based detectors tend to violate this scaling; the baselines of the proposed detectors~\cite{Dimopoulos:2007cj,Hogan:2011tsw,Canuel:2017rrp,Graham:2017pmn,Zhan:2019quq,Tino:2019tkb,AEDGE:2019nxb,Badurina:2019hst,MAGIS-100:2021etm} are typically much shorter than those of a laser interferometer aimed at the same frequency range.} %
with $L \sim L_{\text{opt}} \propto f_{\text{opt}}^{-1}$.
With \eqref{eq:mcmin}, we then see that detectors at higher frequencies will be sensitive to smaller clumps: $\mcl^{\rm min} \propto h^{\text{opt}}_c L_{\text{opt}} \sim f_{\text{opt}}^{-1.7}$.
However, from \eqref{eq:Rate}, we see that this scaling leads to loss in signal rate with increasing frequency: $\dot{N}_{\cl} \propto L_{\text{opt}}^{-1} f_{\text{opt}}^{-2} (h^{\text{opt}}_c)^{-1} \sim f_{\text{opt}}^{-0.3}$. 
Therefore, detectors operating at higher frequencies will typically be sensitive to smaller clump masses and thus complementary to a detector operating in the $\mu$Hz range; however, the expected signal rate (assuming the same clump velocity distribution and mass density) will be lower. 

For a more quantitative comparison of the sensitivities of different existing or proposed detectors to DM clumps, we show their projected sensitivities computed via \eqrefRange{eq:d}{eq:Rate} in \figref{fig:OoM_results}. 
Projections are given for: 
(i) the asteroid-to-asteroid ranging proposal for a baseline of $L \sim 1\,$AU, using the sensitivity curve from \citeR{Fedderke:2021kuy}; 
(ii) a LISA-like detector with $L = 2.5 \times 10^6\,$km, with sensitivity as projected in \citeR{Babak:2021mhe};
(iii) an aLIGO-like detector with $L = 4\,$km and sensitivity taken from \citeR{aLIGO:2018};
(iv) MAGIS-1\,km as an example of an ground-based atom interferometer with $L = 1\,$km and sensitivity based on \citeR{MAGIS-100:2021etm};
and (v) Einstein Telescope (ET) with $L = 10\,$km and strain sensitivity taken from \citeR{Hild:2010id}. 
These projections should be understood as approximate, order of magnitude estimates: e.g., in \figref{fig:OoM_results} we do not account for the velocity distribution one would expect for DM clumps, nor for the orientation of the clump trajectory relative to the detector, nor indeed for the geometry of the TMs in the detector. 
Moreover, the particular detectors or concepts we selected were chosen merely as representative examples spanning a wide range of frequencies and detector technologies; e.g., Cosmic Explorer would have comparable reach to our projections for Einstein Telescope, TianQin's sensitivity would be comparable to that of LISA, and other ground-based atom interferometer proposals~\cite{Canuel:2017rrp,Zhan:2019quq,Badurina:2019hst} have projected sensitivities similar to MAGIS-1\,km. 

A further comment on our sensitivity estimate is in order.
In \figref{fig:OoM_results}, we show the sensitivity projections based on the strain at the peak frequency of the signal, \eqref{eq:fpeak}, by the dashed lines. However, this estimate is too conservative for clumps with $f_{\rm peak} > f_{\rm opt}$ for a given detector; i.e., it underestimates the sensitivity to clumps with smaller masses. 
For simple orientations of the clump trajectory relative to the detector, one can calculate the Fourier transform of the (single TM) acceleration signal analytically (see e.g., \citeR[s]{Seto:2004zu,LISA-Pre-Phase-A}): one finds that the acceleration power spectral density, to which we will return in \sectref{sect:MF}, is approximately \emph{flat} for $f \lesssim f_{\rm peak}$.
Using matched-filter arguments (see \appref{app:matchedFilterScaling}), this observation can be shown to enhance the detectablility of signals with $f_{\rm peak} > f_{\rm opt}$ as compared to the estimate given at \eqref{eq:mcmin}: up to $\mathcal{O}(1)$ factors, the parametric replacement in \eqref{eq:mcmin} is $h_c(f) \rightarrow (\tilde{f}/f) \times h_c(\tilde{f})$, where $\tilde{f}$ denotes the frequency at which $f h_c(f)$ is minimized for a given detector.
This leads to the parametric scaling $\mcl^{\rm min} \propto d$ when $f_{\rm peak}(d) > \tilde{f}$.
In terms of the rate of discoverable signals, this implies $\dot{N}_{\cl} \propto d \propto \mcl$ for $\mcl < \mcl^{\rm min}(f_{\rm peak} = \tilde{f})$.
This improved sensitivity estimate is shown by the solid lines in \figref{fig:OoM_results}.

Let us focus on the sensitivity of the different existing or proposed detectors in terms of the rate of discoverable signals as a function of the clump mass, shown in the right panel of \figref{fig:OoM_results}. 
As we stressed above, we see that detectors operating in different frequency ranges are sensitive to different clump-mass ranges.
The detectors operating at the highest frequencies we consider here are the ground-based laser interferometers aLIGO and ET,%
\footnote{\label{ftnt:tidal}%
    We note that a detectable passage of a less massive object by the higher-frequency, ground-based detector networks such as aLIGO and ET can be required to be so close to the individual detectors in the network that the signal may not appear coincidentally in all the detectors: i.e., maximal values of $d$ can be such that individual transits may be detectable in one detector and not the others in the network.
    Tailored searches for such non-coincident, impulsive events would thus be required.
    Of course, for the regions of parameter space that lie in this `independent transits' limit, overall network event rates would be increased by the number of individual detectors in the network.
    } %
which have the highest rate of discoverable signals for $\mcl \sim 10^2\textup{--}10^3\,$kg; by contrast, the asteroid-ranging concept is the lowest-frequency detector we consider, with largest $\dot{N}_\cl$ to clumps with masses $\mcl \sim 10^{14}\,$kg. 
As we argued above, the scaling of the strain sensitivity of the different detector concepts in the different frequency bands generally leads to the rate of discoverable signals being larger in detectors optimized for lower frequencies~(which are sensitive to heavier clumps). 
Note that the projected strain sensitivity of third-generation ground-based laser interferometers almost overcomes this trend:~the rate of discoverable signals in ET is almost as large as for the asteroid-ranging proposal. 

The scaling of the discoverable rate $\dot{N}_{\cl}$ with $\mcl$ for a given detector is straightforward to understand from the arguments we presented above.
Around the maximum of the $\dot{N}_{\cl}(\mcl)$ curves in the right panel of \figref{fig:OoM_results}, the exact shape of the curve is dictated by the shape of the detector's characteristic-strain sensitivity $h_c(f)$. 
At smaller $\mcl$, where the improved (solid lines in the right panel of \figref{fig:OoM_results}) sensitivity estimate deviates from the overly conservative estimate based on the strain at $f_{\rm peak}$ only (dashed lines), the improved sensitivity estimate scales as $\dot{N}_{\cl} \propto \mcl$, as argued above (see also \appref{app:matchedFilterScaling}). 
For larger clump masses, on the other hand, a detector becomes increasingly sensitive to clumps passing the detector at greater and greater distances, leading to smaller and smaller $f_{\rm peak}$. 
Eventually, the signal will be peaked at frequencies $f_{\text{peak}}(d) \lesssim f_{\text{det}}^{\text{min}}$, where $f_{\text{det}}^{\text{min}}$ is the smallest frequency for which published noised curves are available for a detector.
However, this does not imply complete loss of sensitivity, just a break in the scaling: fixing $d$ to maintain $f_{\text{peak}}(d) \sim f_{\text{det}}^{\text{min}}$ leads to our sensitivity forecast scaling as $\dot{N}_{\cl}(\mcl) \propto 1/\mcl$ owing to the smaller number density of more massive objects; see \eqref{eq:Rate} for fixed $d$. 
This behavior can be seen for the LISA projection at $\mcl \gtrsim 10^{14}\,$kg in the right panel of \figref{fig:OoM_results}; for the other detectors, this transition occurs outside the plotted range of $\dot{N}_{\cl}$.
Finally, we can note that the scaling of $\dot{N}_{\cl}(\mcl)$ for the MAGIS-1km case differs somewhat from the other detectors: MAGIS-1km (and other proposed atom-interferometer detectors) plan(s) to utilize TMs with much smaller separation compared to the (inverse) frequency band for which they are optimized than all other detector concepts for which we show projections in \figref{fig:OoM_results}. 
As a consequence, MAGIS-1km is in the tidal limit for its sensitivity to clumps with $\mcl \gtrsim 100\,$kg, which is below the value $\mcl \sim 10^5\,$kg for which the (non-improved) sensitivity estimate based on the strain at $f_{\rm peak}$ becomes maximal.
This results in scalings $\mcl^{\text{min}} \propto d^2$ and $\dot{N}_{\cl} \propto (\mcl)^0$ between these two masses.
All other detectors for which we show results in \figref{fig:OoM_results} enter the tidal limit only for their sensitivity to clumps with masses larger than those for which we find the largest $\dot{N}_{\cl}(\mcl)$.

It is interesting to compare our results to those of \citeR{Hall:2016usm}, which considered the sensitivity of aLIGO and LISA to the passage of dark clumps using Monte Carlo simulations similar to those that we consider in \sectref[s]{sect:MF} and \ref{sect:FF} for the proposed asteroid-ranging concept detector.
Adjusting for the slightly different assumption for the DM mass density used in \citeR{Hall:2016usm} ($\rho_{\dm}=0.39\,\text{GeV/cm}^3$), Fig.~1 of that reference indicates that aLIGO could detect (at $\rho \geq 1$) $\mcl \sim 10^3\,$kg clumps at a rate of $\dot{N}_{\cl} \sim 3\times 10^{-4}\,\text{yr}^{-1}$, which agrees within an $\mathcal{O}(1)$ factor with the results in our \figref{fig:OoM_results}; this is good agreement given the heuristic nature of our estimates in \figref{fig:OoM_results}.
However, our results for LISA are significantly less optimistic than those of \citeR{Hall:2016usm}: again adjusted for the differing DM mass density, \citeR{Hall:2016usm} projects that clumps of mass $\mcl \sim 10^{11}\,$kg are detectable at $\rho \geq 1$ at a rate of $\dot{N}_{\cl} \sim  8\times 10^{-2}\,\text{yr}^{-1}$, which is a factor of $\sim 25$ \emph{larger} than the rate we project in our \figref{fig:OoM_results}.
We have not been able to resolve this discrepancy definitively.
It is not due to the heuristic nature of our estimate; as we show in \sectref{sect:MF} and \figref{fig:gravPlot}, our heuristic estimation procedure is well validated against Monte Carlo simulations within $\mathcal{O}(3)$ factors at worst.
We speculate that the discrepancy could be due to evolution of LISA design parameters leading us to assume a different projected sensitivity: the preprint for \citeR{Hall:2016usm} appeared before the final LISA~L3 design proposal~\cite{LISA:2017pwj} was released (the strain sensitivity of the latter agrees well with the one we use~\cite{Babak:2021mhe}).

We note that the signal from the passage of an exotic DM clump in a GW detector can be degenerate with the signal from the passage of ordinary objects such as inner Solar System asteroids.
In the non-tidal limit ($d \ll L$), scaling the mass of a transiting object as $m \propto v^2$ and its closest-approach distance as $d \propto v$ yields the same signal waveform for objects transiting at different speeds $v$; see \eqref[s]{eq:d}~and~(\ref{eq:mcmin}).
A closer and less massive ordinary asteroid (slower moving) could thus be mistaken for a more distant and more massive DM clump (faster moving).
This degeneracy is broken when both TMs respond non-trivially to the clump passage (i.e., in the tidal limit).
Of course, in the degenerate case, optical follow-up is possible for an ordinary asteroid, and would allow discrimination of these possibilities~\cite{Seto:2004zu,Adams:2004pk}: nightly all-sky surveys such as ATLAS~\cite{2018PASP..130f4505T} suffice to pick up the larger asteroids that the asteroid-ranging detector would be sensitive to (we estimate this is possible down to $\sim3$\,km asteroid diameters, assuming they are at the closest part of the inner edge of the Main Belt).
Future deep, wide field-of-view images from Rubin Observatory~\cite{2009arXiv0912.0201L} will enable southern-sky optical follow-up for asteroids as small as 150\,m in diameter, at similar distances; likewise, the existing Pan-STARRS instrument~\cite{2016arXiv161205560C,2013PASP..125..357D} continually undertakes wide field-of-view, weekly-cadence, northern-sky monitoring at higher (i.e., fainter) magnitudes than ATLAS, in search of near-Earth asteroids.
Targeted optical follow-up would also be possible. 

The sensitivity of the asteroid-ranging detector to roughly asteroid-mass-scale objects may also be viewed as a science opportunity: because detector sensitivity to $v\sim 10\,$km/s transits (typical for bound objects in the inner Solar System) shifts to less massive objects as compared to DM clumps owing to the speed scalings discussed above, we may be able to use such a detector to augment our knowledge of the incomplete population census and orbital distributions of the smaller, less massive asteroids in the inner Solar System.

Concerning the maximal rate of discoverable signals, we can note from our results in \figref{fig:OoM_results} that, although in particular the asteroid-ranging proposal and next-generation ground-based laser interferometers such as ET come tantalizingly close to being able to discover clumpy DM, the rate of discoverable signals falls just short of what would be required to probe a scenario where all of the DM would be comprised of clumps of a given mass. 
We find the largest rate of discoverable signals out of all detectors for the asteroid-ranging concept, $\dot{N}_{\cl} \sim 0.05\,{\rm yr}^{-1}$ for $\mcl \sim 10^{14}\,$kg. 
Such a rate corresponds to one discoverable signal during a $\sim 20\,$yr mission. 
The rate of discoverable signals could be larger than these projections if, for example, the clump density in the vicinity of the Solar System would be much larger than our assumption of $\rho_{\cl}\sim \rho_{\dm} \sim 0.3\,$GeV/cm$^3$. 
Current measurements of the ``local'' DM density are sensitive to the average DM density over volumes larger than $\sim (100\,{\rm pc})^3$, leaving open the possibility that the DM density in the close vicinity of the Solar System is much larger than $0.3\,$GeV/cm$^3$; for instance, we could be in a DM stream or other over-dense sub-structure~\cite{Moore:1999nt,Diemand:2008in}.

Another possibility for a larger rate of discoverable signals is if there is an additional long-ranged force between the clumps and ordinary matter (including the detector TMs), a possibility which we will entertain further in Sec.~\ref{sect:FF}. 
If, on the other hand, we maintain our assumption of $\rho_{\cl} \sim \rho_{\dm} \sim 0.3\,$GeV/cm$^3$ and purely gravitational interactions of the DM clumps with the detector, we can read off the improvement in sensitivity that would be required for any given detector concept to achieve any desired rate of discoverable signals. 
Keeping the baseline $L$ of a detector fixed but improving the characteristic-strain sensitivity by a factor $x$, would shift the corresponding $\dot{N}_{\cl}(\mcl)$ curve in the right panel of \figref{fig:OoM_results} upwards in $\dot{N}_{\cl}$ and leftwards to lower $\mcl$ by the same factor $x$, as can be seen from equations \eqref[s]{eq:mcmin} and (\ref{eq:Rate}). 
For example, an asteroid-ranging detector with strain sensitivity improved by a factor of $\sim 20$ compared to the projection in Ref.~\cite{Fedderke:2021kuy} could detect one event per year if all of the DM is comprised of $\mcl \sim 10^{13}\,$kg clumps, while a ground-based laser interferometer with $L \sim 10\,$km and a characteristic strain sensitivity $\sim 50$ times better than the projections for ET/CE could detect one event per year if all of the DM is made up of $\sim 1\,$kg clumps.

Finally, it is also worth commenting on PTA sensitivity, as these detectors are in a different class to those with all their TMs located within the Solar System.
PTA searches for GWs gain an advantage in their strain sensitivity by virtue of being able to consider effective baselines up to the GW wavelength, $\lambda_{\gw} \sim 2\times 10^3\,\text{AU} \times ( \mu\text{Hz}/f_{\gw})$.
However, as a result of this, the sensitivity of a PTA to accelerations of the Earth or pulsar (accessed observationally via the Doppler effect~\cite{Seto:2007kj,Dror:2019twh,Ramani:2020hdo}) is, for a given GW strain sensitivity, much poorer than that of a local-TM-based detector with the same strain sensitivity: if the local-TM-detector has a baseline $L$, it will have a factor of $\sim \lambda_{\gw}/L \gg 1$ better acceleration sensitivity than the PTA.
As such, PTAs are generally not well-suited for searching for impulsive acceleration signals from asteroid-mass-scale DM transits, where the baseline length is irrelevant to the size of the signal (modulo the possible tidal nature of the acceleration).
PTAs do however have complementary sensitivity to accelerations caused by transiting objects somewhat more massive than we consider in this work~\cite{Seto:2007kj,Dror:2019twh,Ramani:2020hdo}.
There is also a second signal that PTAs are sensitive to: pulsar timing shifts caused by Shapiro delays due to DM transits close to the Earth--pulsar line of sight. 
These are not as severely suppressed as might na\"ively be expected, owing to the large distance to the pulsars~\cite{Dror:2019twh}; however, detectable signals typically require even larger DM masses than the Doppler-based searches~\cite{Siegel:2007fz,Dror:2019twh,Ramani:2020hdo,Schutz:2016khr}.

\section{Matched-Filter Search}
\label{sect:MF}

So far, we have presented a simplified estimate of the sensitivity of GW detectors to the passage of a clump. 
In this section, we present a more careful estimate of the sensitivity based on a matched-filter search for the signal induced by a clump.

Since the precise form of the signal in a GW detector depends on the detector geometry, we will show quantitative results for only the particular case of ranging the distance between two asteroid test masses by means of a direct laser/radar ranging link, as proposed in \citeR{Fedderke:2021kuy}.
Our formalism is straightforward to extend to more complicated detector geometries. 
Furthermore, we will account for the geometry of the clump trajectories relative to the detector expected for a given velocity distribution of the clumps, as well as gravitational effects on the clump trajectories during their passage through the gravitational potential of the Sun. 
We will show numerical results for the particular case of the clump velocity distribution following the Standard Halo Model~\cite{Drukier:1986tm,Lewin:1995rx}; i.e., a Maxwell--Boltzmann distribution truncated at the galactic escape speed and boosted into the Solar System frame.
We will assume two possible TM configurations: a toy example of ranging two (fictitious) asteroids located 0.5\,AU from the Sun, separated from each other by 1\,AU, and fixed in space (``Sun-straddling 1\,AU baseline''); and a more realistic setup of ranging between two known asteroids with appropriate properties to serve as good TMs, \Eros\ and \Thomsen. 
The formalism we present readily extends to any other velocity distribution and locations of the TMs.

Let us begin by writing \eqref{eq:acc}, the acceleration a clump exerts on a TM, more carefully. 
If we denote the location of the $i$-th TM (in any given coordinate system) as $\bm{r}_i$ and the location of the clump as $\bm{r}_{\cl}$, the gravitational acceleration of the $i$-th test mass is
\begin{align} \label{eq:acc2}
    \bm{a}_i(t) = G_N \mcl \frac{\bm{r}_{\cl}(t) - \bm{r}_i(t)}{\left| \bm{r}_{\cl}(t) - \bm{r}_i(t) \right|^3} \:.
\end{align}
A GW detector is sensitive to the relative acceleration of two TMs projected onto the baseline,
\begin{align} \label{eq:BPDA}
    \Delta a_{ij}(t) \equiv \left[ \bm{a}_i(t) - \bm{a}_j(t) \right] \cdot \left[ \frac{\bm{r}_i(t) - \bm{r}_j(t)}{\left| \bm{r}_i(t) - \bm{r}_j(t) \right| } \right] \:.
\end{align}
This expression readily extends to multiple clumps accelerating the TMs at the same time; to account for this we simply need to replace $\bm{a}_i \to \bm{a}_i^{\rm tot} \equiv \sum_k \bm{a}_i^k(t)$, where $\bm{a}_i^k(t)$ is the acceleration of the $i$-th test mass by the $k$-th clump given by \eqref{eq:acc2} [with $\bm{r}_{\cl} \to \bm{r}_{\cl}^k$]. 

In order to perform a matched-filter search, one computes the Fourier transform of $\Delta a_{ij}(t)$,
\begin{align} \label{eq:FT}
    \widetilde{\Delta a}_{ij}(f) \equiv \int dt\,e^{2\pi i f t} \Delta a_{ij}(t) \:.
\end{align}
The signal-to-noise ratio (SNR) $\rho$ is then~\cite{Maggiore:2007zz,Moore:2014lga,Fedderke:2020yfy}
\begin{align} \label{eq:SNR}
    \rho^2 \equiv 4 \int\limits_0^\infty df\, \frac{\left| \widetilde{\Delta a}_{ij}(f) \right|^2 }{S_a(f)} \:,
\end{align}
where $S_a(f)$ is the (one-sided) power spectral density (PSD) of the relative (baseline-projected) acceleration noise at the frequency $f$. 
For the purpose of this work, we estimate $S_a(f)$ from published strain noise curves. These are typically published in terms of either the PSD noise of the strain, $S_h(f)$, or the characteristic strain sensitivity, $h_c(f)$, which are related via $h_c(f) \equiv \sqrt{f S_h(f)}$; see, e.g., \citeR{Moore:2014lga}. 
From $S_h$ (or $h_c$) we estimate $S_a$ as~\cite{Fedderke:2021kuy}
\begin{align}
    S_a(f) \sim \frac{1}{2} \left( 2 \pi f \right)^4 L^2 S_h = \frac{1}{2} \left( 2 \pi \right)^4 f^3 L^2 h_c^2(f) \:,
    \label{eq:SaRelationship}
\end{align}
where $L$ is the length of the detector baseline.

The remaining task to evaluate the sensitivity of a GW detector to the passage of a clump is to compute the trajectories of the clumps in the vicinity of the TMs; i.e., for our purposes of a detector with TMs in the inner Solar System, to compute $\bm{r}_\cl(t)$ through the inner Solar System.%
\footnote{\label{ftnt:discreteFT}%
    Note that technically we compute the $\bm{r}_{\cl}^k(t)$, and, in turn, $\Delta a_{ij}(t)$, as discrete time series.
    The Fourier transform in \eqref{eq:FT} is accordingly replaced by a discrete Fourier transform and the integral in \eqref{eq:SNR} by a sum over discrete frequencies.
    See, e.g., \citeR{Fedderke:2020yfy} for detailed formulae. } %
To this end, we perform a Monte Carlo simulation to draw initial conditions for clumps with a homogeneous density $\rho_{\cl}$ and velocity distribution $f(\bm{v})$ entering a sphere with radius $R_0$ centered on the Sun [the ``trajectory-initialization (TI) sphere'']. 
We evolve the orbits from these initial conditions in the (Newtonian) gravitational potential of the Sun. 
For our numerical results, we use a velocity distribution of the clumps following the Standard Halo Model (SHM); i.e., a Maxwell--Boltzmann distribution with velocity dispersion $\sigma_v = 156\,$km/s in the Milky Way (MW) galactic rest frame, truncated at the galactic escape velocity $v_{\rm esc} = 544\,$km/s, and boosted by the Solar System's speed $v_\odot = 220\,$km/s toward Cygnus in the galactic rest frame~\cite{Evans:2018bqy}.
We normalize the flux of clumps through the TI sphere to $\rho_{\cl} = \kappa \times 0.3\,$GeV/cm$^3$, where $\kappa$ is the fraction of the DM mass density in clumps.
For the purposes of this section, we assume $\kappa = 1$.
More discussion of the technical details of our calculation can be found in \appref{app:technicalDetail}.

\begin{figure*}
    \centering
    \includegraphics[width=\textwidth]{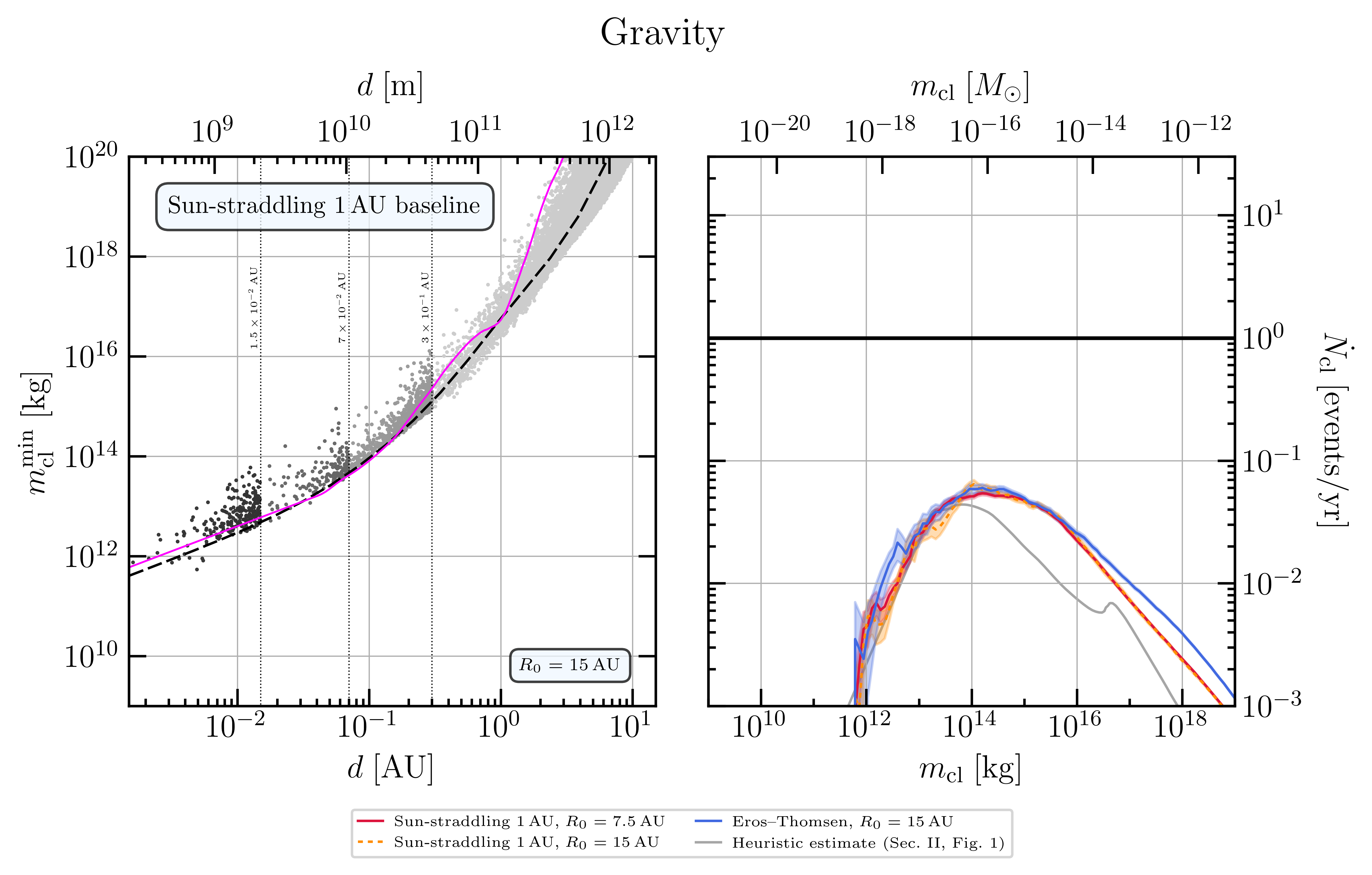}
    \caption{%
    Sensitivity projections based on our matched-filter search for an asteroid-to-asteroid ranging GW detector~\cite{Fedderke:2021kuy}, assuming purely gravitational interactions between the clumps and ordinary matter (including the detector TMs). 
    \textsc{Left panel:}~Smallest clump mass $\mcl^{\text{min}}$ leading to a detectable signal (defined as having a SNR of $\rho \geq 1$) as a function of the smallest achieved distance $d$ of the clump to either of the test masses. 
    These results assume the ``Sun-straddling 1\,AU baseline'' TM configuration. 
    Each point in the scatter plot corresponds to a trajectory from a Monte Carlo simulation with a TI sphere of radius $R_0 = 15\,$AU, assuming a clump velocity distribution given by the Standard Halo Model. 
    The different gray-shades of the clumps delineated by the vertical dotted black lines correspond to different disjoint parts of our simulation; see the text, and in particular \appref{app:rateExtraction}, for a technical discussion including the merging of the different simulations. 
    For comparison, the dashed black line shows the matched-filter results for the specific `selected' trajectory orientation discussed in the text, and the solid magenta line shows the (improved) heuristic estimate from Sec.~\ref{sect:OoM}. 
    \textsc{Right panel:}~Rate of discoverable ($\rho\geq1$) signals  $\dot{N}_{\cl}$ assuming that clumps of mass $\mcl$ have a local density of $\rho_{\cl} = 0.3\,$GeV/cm$^3$. 
    The different colored lines show results for different  choices of the simulation setup and the detector TM configuration as denoted in the legend and discussed in the text. 
    The shaded band around each colored line gives the $1\sigma$ Poisson error bands for our Monte Carlo simulation. 
    The solid gray line shows the heuristic estimate from Sec.~\ref{sect:OoM} (\figref{fig:OoM_results}). 
    }
    \label{fig:gravPlot}
\end{figure*}

In \figref{fig:gravPlot} we show sensitivity projections based on the matched-filter search.
Throughout this work, we consider a signal detectable if the SNR satisfies $\rho \geq 1$.
In the left panel of \figref{fig:gravPlot} we show the minimal clump mass required to give rise to a detectable signal as a function of the smallest achieved distance of the clump to either of the TMs, $\mcl^{\rm min}(d)$, analogous to the left panel of \figref{fig:OoM_results}. 
Each of the points in this scatter plot corresponds to a simulated trajectory. 
The spread in points in the vertical direction (i.e., the range of $\mcl^{\rm min}$ we obtain for a given $d$) is due to the distribution of trajectories: for a given $d$, clumps pass the detector with different relative speeds and with different orientations of their trajectories with respect to the detector baseline.
For computational efficiency, the computation is split into different intervals in $d$ as indicated by the thin dotted vertical lines and the different shades of points from different simulations. 
The results in this panel assume the fictitious ``Sun-straddling 1\,AU baseline'' configuration (two asteroids located 0.5\,AU from the Sun, separated from each other by 1\,AU, and fixed in space), and we evolve the orbits in the gravitational potential of the Sun starting from a TI sphere with $R_0 = 15\,$AU. 
For comparison, the dashed black line shows the matched-filter $\mcl^{\rm min}(d)$ relation we find for a particular encounter geometry: choosing a Cartesian coordinate system where the TMs are located at $\bm{r}_i = (\pm L/2,0,0)$, we consider a clump passing the detector with velocity $\bm{v}_{\cl} = (0,0,\vcl)\,$ and point of closest approach $\bm{r}_{\cl}=(L/2+d,0,0)$. We compute $\vcl$ from taking into account the gravitational effects of the Sun on a clump with asymptotic speed $\vcl^\infty = 300\,$km/s relative to the Solar System. 
The solid magenta line shows the (improved) heuristic estimate from Sec.~\ref{sect:OoM}.

In the left panel of \figref{fig:gravPlot}, the heuristic estimate, the result for the particular geometry, and the scatter plots showing the results from our Monte Carlo simulation, all agree rather well. 
This demonstrates first that the heuristic estimate gives us a good understanding of the more careful projection based on a matched-filter search, and second, that our Monte Carlo simulation reproduces the results from a simpler matched-filter calculation using a fixed geometry, up to expected spread from different possible clump trajectories for a given $d$. 
Taking a closer look, one can note that the heuristic estimate is overly conservative for clumps passing the TMs with $d \gtrsim 0.1\,$AU, or in terms of the clump mass, for $\mcl^{\rm min} \gtrsim 10^{14}\,$kg. 
As discussed in Sec.~\ref{sect:OoM}, for $d \gtrsim 0.1\,$AU the strain signal is peaked at frequencies $f_{\rm peak}$ \emph{smaller} than the ``corner frequency'' $f_{\rm cnr}$ at which the detector sensitivity curve starts degrading faster than $h_c \propto f^{-2}$. 
The reason that the matched-filter search yields better sensitivity for $d \gtrsim 0.1\,$AU than our heuristic estimate is that the signal $\widetilde{\Delta a}_{ij}(f)$, as given by \eqref{eq:FT}, falls off slower at frequencies larger than $f_{\rm peak}$ than does the detector noise $S_a(f)$ in this regime: tail effects not captured by the heuristic estimate boost the sensitivity.
Modeling this effect in our heuristic estimate would have required detailed modeling of the frequency content of the signal and the shape of the detector noise curve; the matched-filter search by design accounts for such effects.

In the right panel of \figref{fig:gravPlot} we show projections for the rate of detectable events we expect as a function of $\mcl$, analogous to the right panel of \figref{fig:OoM_results}. 
We show results from our Monte Carlo simulations for three cases: the solid red lines show results for the ``Sun-straddling 1\,AU'' baseline configuration with orbits evolved from a $R_0 = 7.5\,$AU TI sphere, while the dashed orange lines show results for the same TM configuration but for $R_0 = 15\,$AU; finally, the solid blue lines show projections for TMs that follow (osculating) elliptical orbits corresponding to the known asteroids \Eros~and \Thomsen~(see \tabref{tab:ErosThomsenOrbitalElements} in \appref{app:rateExtraction} for details) and $R_0 = 15\,$AU. 
For comparison, the gray line shows the projections from our (improved) heuristic estimate from Sec.~\ref{sect:OoM} (see \figref{fig:OoM_results}). 
For the Eros--Thomsen case, the TMs move, and we have to make assumptions about the duration and time of the observations; we use a $T=10$\,yr mission duration centered on one of the epochs of perihelion passage of \Eros, and all rate results should be understood to be averaged over that duration.

Comparing the matched-filter projections with the heuristic estimate, we find good agreement for $\mcl \lesssim 10^{14}\,$kg; for larger masses, the heuristic estimate of detectable events is overly conservative. 
This is due to the reasons discussed in the previous paragraph; i.e., in this regime, the heuristic estimate does not properly account for the frequency content of the signal at $f \gtrsim f_{\rm peak}$. 
Looking at the matched-filter projections for the three different cases shown in the right panel of \figref{fig:gravPlot}, we can first note that the ``Sun-straddling 1\,AU'' baseline configuration results for the two different choices of $R_0$ agree excellently with each other. 
This validates our sampling of the initial conditions at the TI sphere, a non-trivial exercise discussed in \appref{app:technicalDetail}. 
Comparing the results for this fictitious TM configuration with the projections for the Eros--Thomsen TM case, we find that the latter TM configuration performs slightly better. 
We caution that these results should be taken with a grain of salt; in the absence of dedicated sensitivity projections for particular TMs (and baseline configurations), we use the same sensitivity curve from Ref.~\cite{Fedderke:2021kuy} for both TM configurations. 
Thus, the only difference between these two TM configurations is that for the Eros--Thomsen case, the TMs (and hence the location, length, and orientation of the baseline) changes with time, while for the ``Sun-straddling 1\,AU'' case, the TMs are fixed in space and the baseline has a fixed length of $1\,$AU. 
The moving TMs and, in particular, the different lengths of the baselines, lead to small differences in the sensitivity projections. 
The most pronounced difference is at $\mcl \gtrsim 10^{16}\,$kg.
Consulting the left panel of \figref{fig:gravPlot}, we can note that clump masses $\mcl \gtrsim 10^{16}\,$kg are still detectable for $d \gtrsim 1\,$AU; i.e., trajectories for which the interaction with the ``Sun-straddling 1\,AU'' detector enters the tidal limit. 
For the \Eros\ and \Thomsen\ orbits, the length of the baseline changes (slowly) with time, $1.1\,{\rm AU} \lesssim L \lesssim 4.6\,$AU; the time-averaged length of the baseline is $\bar{L} \sim 3.1$\,AU. 
Due to the longer baseline, the transition from the single-TM-acceleration limit to the tidal limit of the acceleration signal from the passage of a clump occurs at somewhat larger $d$, leading to improved sensitivity to clumps with $\mcl \gtrsim 10^{16}\,$kg.
Although we do not show the results here, the large-$\mcl$ limit of the Eros--Thomsen results agrees well with simulation results for a ``Sun-straddling 3\,AU'' baseline.

Returning to broader considerations of the sensitivity projections, we repeat our conclusions from Sec.~\ref{sect:OoM}: 
Although the asteroid-ranging proposal comes tantalizingly close to being able to discover clumpy DM, the rate of discoverable signals falls just short of what would be required to probe a scenario where all of the DM would be comprised of clumps of a given mass. 
We find a largest rate of discoverable signals of $\dot{N}_{\cl} \sim 0.05\,{\rm yr}^{-1}$ for $\mcl \sim 10^{14}\,$kg, or one discoverable signal during a $\sim 20\,$yr mission. 
These projections are based on a clump density of $\rho_{\cl}\sim \rho_{\dm} \sim 0.3\,$GeV/cm$^3$, a clump velocity distribution as given by the SHM, and purely gravitational interactions of the clumps with the TMs. Larger signal rates are possible if the local DM density in the close vicinity of the Solar System is larger or if the clumps follow a different velocity distribution. 
Another possibility leading to a larger rate of discoverable signals is an additional long-ranged force between the clumps and ordinary matter (including the detector TMs), which we will entertain in the following section.

\section{Fifth Force}
\label{sect:FF}
So far, we have considered the sensitivity of GW detectors to the nearby passage of clumps that interact purely gravitationally with ordinary matter such as the detector TMs or the Sun. 
In this section, we entertain the possibility that the clumps are comprised of a form of matter which is subject to an additional attractive long-range force; see also \citeR{Hall:2016usm} for a similar study for LISA and aLIGO.

There are strong experimental limits on new (long-range) forces between ordinary (Standard Model, SM) particles from, e.g., precision tests of the gravitational law ranging from the micrometer to the few-AU scale and null-results of searches for violations of the equivalence principle; see, e.g., \citeR[s]{Adelberger:2009zz,Zyla:2020zbs}. 
However, even if the clumps make up all of the DM, constraints on (long-range) forces between SM matter and the clumps, or clump-clump forces, are much weaker; moreover, any such limits disappear entirely when considering the clumps to make up only a small fraction of the DM, conservatively $\kappa \lesssim 1\,\%$.

\begin{figure}
    \centering
    \includegraphics[width=\columnwidth]{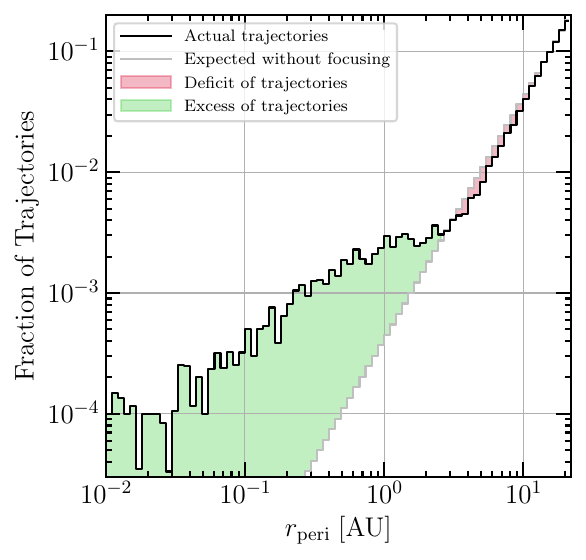}
    \caption{The black line shows the perihelion distribution of clump trajectories from our Monte Carlo simulation for a stronger-than-gravity ($\xi =10^3$) SM--DM fifth force with range $\lambda = 3\,$AU, for initial conditions appropriate to a SHM velocity distribution of the clumps at a TI sphere of radius $R_0 = 22\,$AU. 
    For comparison, the gray line shows the expected perihelion distribution without a focusing force. 
    The shaded green (respectively, red) region highlights the excess (deficit) of trajectories with perihelia $r_{\rm peri} \lesssim \lambda$ ($r_{\rm peri} \gtrsim \lambda$) compared to the no-focusing case. 
    Note that the logarithmic scaling of the axes obscures that the total number of trajectories is conserved (i.e., the red and green shaded areas represent the same number of trajectories). 
    }
    \label{fig:perihelionPlot}
\end{figure}

We will consider a new force between a clump with mass $\mcl$ and a SM object with mass $M$ that gives rise to an attractive Yukawa potential
\begin{align}
    V_{5'}(r) = - \xi \frac{G_N \mcl M}{r} e^{-m_\phi r} \:,
\end{align}
where $r$ denotes the distance between the clump and the ordinary object, $m_\phi$ is the mass of the particle mediating the new force, and $\xi$ parameterizes the strength of the force relative to gravity (at $r \ll 1/m_\phi$). 
This new force will have two important effects on the signal such a clump would induce in a GW detector when passing through the Solar System.
First, the acceleration of the detector TMs would no longer be given by \eqref{eq:acc2}; instead,
\begin{align} \label{eq:accFF}
    \bm{a}_i(t) = G_N \mcl \frac{\bm{r}_{\cl, i}}{ r_{\cl, i}^3} \left[ 1 + \xi \left( 1 + m_\phi r_{\cl, i} \right) e^{- m_\phi r_{\cl, i} } \right] \:,
\end{align}
where $\bm{r}_{\cl, i} \equiv \bm{r}_{\cl}(t) - \bm{r}_i(t)$ and $r_{\cl, i} = |\bm{r}_{\cl, i}|$. 
Second, this new force will modify the clump trajectories through the Solar System.
In our Monte Carlo simulation of the trajectories, we take into account the effect of the Sun on the clumps via both the new interaction and gravity; we will be interested in long range forces with range $\lambda \equiv 1/m_\phi \gtrsim 1\,$AU, such that the effect of the Sun--clump interaction will dominate in the inner Solar System and we thus we ignore the effects of the planets and other bodies on the clump trajectories.
The effects of the new force on the clump trajectories are again twofold: the additional force
(a) accelerates clumps as they fall into the potential of the new force, and 
(b) focuses the trajectories. 
In order to illustrate the latter effect, we show in \figref{fig:perihelionPlot} the \emph{perihelion} distribution (i.e., the distribution of the closest approaches of the clump trajectories to the Sun) for a fifth force with range $\lambda = 3\,$AU and strength $\xi = 10^3$. 
The black line shows the histogram of the trajectories obtained from our Monte Carlo simulation for orbits with a SHM velocity distribution at a TI sphere of radius $R_0 = 22\,$AU.
The gray line shows the distribution one would obtain without focusing, and the red- and green-shaded areas highlight the focusing effect of the fifth force.
Note that in the range shown in \figref{fig:perihelionPlot}, the perihelion distribution for the same initial conditions but with gravity-only interactions (as in Sec.~\ref{sect:MF}) would look indistinguishable (up to statistical fluctuations) from the expectation without focusing.

Before we turn to our sensitivity projections for the asteroid-to-asteroid ranging proposal to clumps with this additional fifth force, let us sketch how this force could arise such that limits on additional forces between SM particles and on DM self-interactions are satisfied. 
Consider clumps composed of new constituent particles $\chi$ with mass $m_\chi$, and a new force mediated by a light mediator $\phi$,
\begin{align} \label{eq:LFF}
    \mathcal{L} \supset -g \phi \bar{N} N - g' \phi \bar{\chi} \chi + \frac{1}{2} m_\phi^2 \phi^2 \:.
\end{align}
We can then parameterize the new force between ordinary nucleons $N$ by $\alpha_5 = g^2/4\pi$, the $\chi$--$\chi$ coupling by $\alpha_D^2 = (g')^2/4\pi$, and the SM--$\chi$ coupling by $\alpha_{5'} = g g'/4\pi$. 
To connect to our notation above, we identify $\xi = (\alpha_{5'}/\alpha_G) \times (m_p/m_\chi)$, where we choose $\alpha_G \equiv G_N m_p^2$ to parameterize the strength of gravity using the gravitational coupling between protons with mass $m_p$.
Similarly, we can measure the strength of the new SM--SM force relative to the strength of gravity with $\epsilon \equiv \alpha_5/\alpha_G$. 
Precision tests of gravity in the Solar System and in the laboratory constrain%
\footnote{Here, we assume that the new force couples to the SM proportional to baryon number as suggested in \eqref{eq:LFF}. For couplings to either protons or neutrons only, the constraint is $\epsilon \lesssim 10^{-11}$.} %
$\epsilon \lesssim 10^{-10}$ in the force range we are most interested in, $\lambda = 1/m_\phi \sim 1\,$AU~\cite{Adelberger:2009zz}. 
Numerically, $\alpha_G \sim 10^{-38}$, such that one can achieve a stronger-than-gravity SM--clump force (i.e., $\xi > 1$) with tiny $\alpha_D$ while satisfying these bounds on $\epsilon$: as long as $\alpha_D > (m_\chi/m_p)^2 \times (\alpha_G/\epsilon)$, the condition $\xi > 1$ is satisfied. 
For example, taking $\epsilon = 10^{-10}$ and $m_\chi = m_p$, a dark-sector coupling of $\alpha_D \sim 10^{-28}$ leads to $\xi \sim 1$. 

If the clumps constitute all of the DM, one might be concerned that the new clump--clump force leads to strong DM self-interactions. 
Indeed, the clump--clump total scattering cross section is enormous, $\sigma \sim \alpha_D^2 \mcl^2 / m_\phi^4$. 
However, this cross section is dominated by very soft interactions with typical momentum transfer $\Delta q \sim m_\phi$. 
Limits on DM self-interactions mostly constrain the so-called momentum-transfer cross section~\cite{Feng:2009hw,Buckley:2009in,Tulin:2013teo,Yang:2022hkm}, which, for this model and for DM velocities $v_{\rm DM} \gg m_\phi/\mcl$, is $\sigma_T \sim \alpha_D^2 / \mcl^2 v_{\rm DM}^4$. 
Limits on the DM self-interaction cross sections are roughly $\sigma_T/m_\cl \lesssim 1\,{\rm cm^2}/$g at $v_{\rm DM} \sim 1000\,$km/s; this bound is satisfied in this model by an enormous margin:
\begin{align}
\begin{split}
    \frac{\sigma_T}{\mcl} \sim 10^{-172}\,\frac{{\rm cm}^2}{\rm g} &\times \left( \frac{\alpha_D}{10^{-28}} \right)^2 \times \left( \frac{10^{14}\,{\rm kg}}{\mcl} \right)^3 \\
    & \times \left( \frac{1000\,{\rm km/s}}{v_{\rm DM}} \right)^4 .  
\end{split} 
\end{align}
A detailed consideration of the bounds on, and effects of, such a new force on, e.g., clump--clump dynamics is beyond the scope of this work. 
The arguments above are advanced merely to illustrate how a stronger-than-gravity fifth force between the clumps and ordinary matter could arise without violating existing constraints.
As we will see, the asteroid-to-asteroid ranging proposal could be sensitive to the passage of clumps which have such additional interactions even if the clumps have a mass density much smaller than that of DM; we will consider $\rho_{\rm cl} = \kappa \times 0.3\,$GeV/cm$^3$ with $\kappa = 10^{-2}$.
In this case, DM self-interaction constraints disappear.

\begin{figure*}
    \centering
    \includegraphics[width=\textwidth]{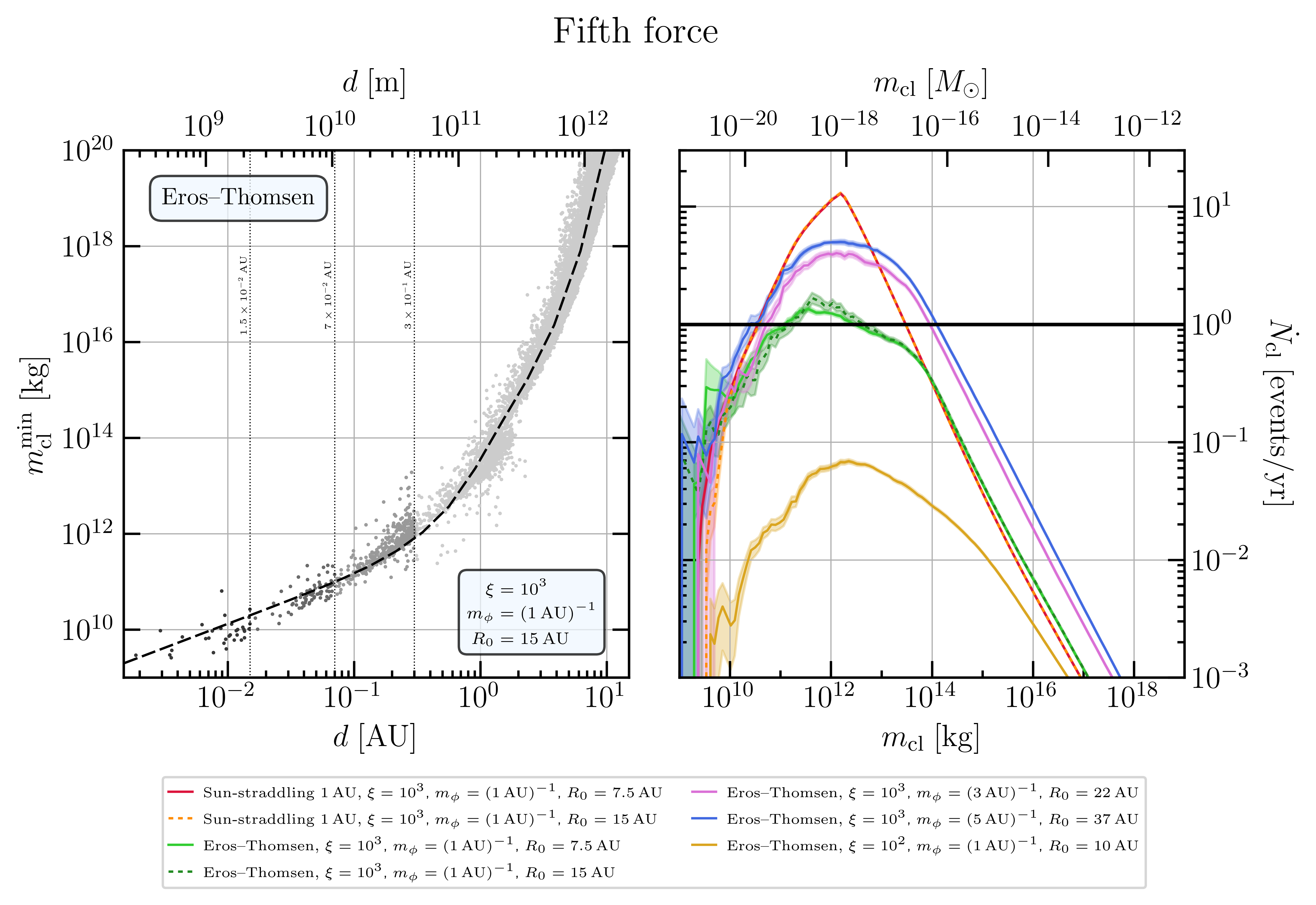}
    \caption{%
    Same as \figref{fig:gravPlot} except that we include an attractive, stronger-than-gravity SM--clump fifth force.
    \textsc{Left panel:} We assume a fifth force  stronger than gravity ($\xi = 10^3$) and with range $1/m_\phi = 1\,$AU. 
    The scatter plot shows the $\mcl^{\rm min}(d)$ results for trajectories from our Monte Carlo simulations of clump trajectories under the influence of Solar gravity and the Sun-sourced fifth force, starting from a TI sphere with $R_0 = 15\,$AU, assuming the Eros-Thomsen TM configuration, and accounting for the clump--TM fifth force.
    For comparison, the black dashed line shows $\mcl^{\rm min}(d)$ for the same specific `selected' trajectory as in \figref{fig:gravPlot}, except that we account for the effects of the fifth force; note that this line assumes the ``Sun-straddling 1\,AU baseline'' configuration and so is not a perfect benchmark.
    \textsc{Right panel}: Rate of discoverable signals as a function of the clump mass, $\dot{N}_{\cl}(\mcl)$, for different choices of the TM configuration, the strength and range of the fifth force, and the radius of the TI sphere as denoted in the legend.
    For all results, we assume that the clumps have a density (at the TI sphere) corresponding to $\kappa = 10^{-2}$ of the DM, $\rho_{\cl} = \kappa \times 0.3\,$GeV/cm$^3$. 
    Projections for different values of $\kappa$ are trivially obtained by rescaling $\dot{N}_{\cl}(\mcl) \propto \kappa/10^{-2}$.
    }
    \label{fig:FFPlot}
\end{figure*}

We now turn to the sensitivity projections for the asteroid-to-asteroid ranging proposal to the passage of clumps under the assumption of a new SM-clump force;
these are shown in \figref{fig:FFPlot}.
We perform a Monte Carlo simulation of orbits drawn at a heliocentric TI sphere with radius $R_0$ as in \sectref{sect:MF}.
Similar to how we included Sun--clump gravitational effects on the clump trajectories in \sectref{sect:MF}, we now include the effects of the fifth-force potential sourced by the Sun on the clump trajectories. 
As in \sectref{sect:MF}, we assume that the velocity distribution of the clumps at the TI sphere is given by the SHM. 

The left panel of \figref{fig:FFPlot} shows the smallest clump mass giving rise to a detectable signal as a function of the smallest separation distance of the clump trajectory to either of the detector TMs, analogous to the left panels of \figref[s]{fig:OoM_results} and~\ref{fig:gravPlot}. 
We set the strength of the fifth force to $\xi = 10^{3}$, assume a range of $\lambda = 1/m_\phi = 1\,$AU, and assume the Eros--Thomsen TM configuration.%
\footnote{%
    Note that since we are showing $\mcl^{\rm min}$ as a function of the smallest distance along a trajectory to either TM, $d$, these results are largely independent of the TM configuration. 
    For different TM configurations, small differences would arise in the $d$-distribution of the points in the scatter plot: each point corresponds to an orbit from our simulation which is subject to the focusing effect of the Sun. 
    The velocities of the clumps in the vicinity of the TMs will differ slightly due to the different TM--Sun distances for different TM configurations, leading to minor differences in $\mcl^{\rm min}$ for a given $d$.} %
Comparing to the gravity-only result in the left panel of \figref{fig:gravPlot}, we find that, due to the fifth force, the mass of a clump required to give rise to a detectable signal is approximately a factor of $\xi = 10^3$ smaller at $d \lesssim \lambda$, while for $d$ large compared to the range of the fifth force, $\mcl^{\rm min}$ converges to the gravity-only result. 

In the right panel of \figref{fig:FFPlot} we show the expected rate of discoverable events for a range of different assumptions on the strength and range of the fifth force and the configuration of the TMs, as denoted in the legend. 
First, note that compared to the gravity-only results in the right panel of \figref{fig:gravPlot}, we find much larger detectable event rates for our choices of the strength ($\xi = 10^3$, except for one case with $\xi = 10^2$) and the range of the fifth force ($1/m_\phi = 1\textup{--}5\,$AU), despite assuming that the clumps have mass density (at the TI sphere) corresponding to only a small fraction of the DM, $\rho_\cl = \kappa \times 0.3\,$GeV/cm$^3$ with $\kappa = 10^{-2}$.
Furthermore, we now find the peak of the $\dot{N}_{\cl}(\mcl)$ curves at smaller clump masses, $\mcl \sim 10^{12}\,$kg, than for the gravity-only case where we found the largest $\dot{N}_{\cl}$ at $\mcl \sim 10^{14}\,$kg. 
Both effects are easy to understand. 
Due to the additional clump--TM interaction, a clump passing the TM at a fixed distance now exerts a stronger force on the TM; or, by the same token, a detectable acceleration signal can still be obtained for a smaller clump mass. 
Due to the effects of the additional clump--Sun interactions, orbits are focused in the inner Solar System, further enhancing the encounter rate. 
These effects are stronger than the suppression of the rate of discoverable signals from assuming a smaller clump density, $\kappa \sim 10^{-2}$.
Note that the results shown in the right panels of \figref[s]{fig:OoM_results},~\ref{fig:gravPlot}, and~\ref{fig:FFPlot} can be trivially rescaled to obtain results for any other value of $\rho_{cl}$ (or $\kappa$): $\dot{N}_{\cl} \propto \kappa/\kappa_{\rm ref}$ where $\kappa_{\rm ref}$ is the choice made in the corresponding plot.

Let us now discuss the results in the right panel of \figref{fig:FFPlot} in more detail, which will allow us to disentangle the different effects of the fifth force. For the ``Sun-straddling 1\,AU baseline'' TM configuration and for the Eros-Thomsen case, we show two sets of results for the same assumptions on the strength of the fifth force ($\xi = 10^3$) and its range ($1/m_\phi = 1\,$AU), in solid and dashed lines. 
The difference between these results for a given TM configuration is solely the choice of the radius of the TI sphere: $R_0 = 7.5$\,AU for the solid lines and $R_0 = 15\,$AU for the dashed lines. 
The results for different $R_0$ are in excellent agreement, again validating our procedure for sampling the initial conditions for the orbits at the TI sphere. 
Note that we choose $R_0$ such that $\xi e^{-m_{\phi} R_0} < 1$; i.e., the Sun--clump potential is suppressed to that of gravity at $R_0$.  

Comparing the $\xi = 10^3$, $1/m_\phi = 1\,$AU results for the ``Sun-straddling 1\,AU baseline'' TM configuration with the Eros-Thomsen case, we find that for $\mcl \lesssim 10^{14}\,$kg, $\dot{N}_{\cl}$ is much smaller for the Eros-Thomsen TM configuration. 
This is due to the focusing effect of the fifth force on the clump trajectories. 
Consulting the left panel of \figref{fig:FFPlot}, we see that for these choices of fifth force parameters, clumps are detectable for $\mcl \lesssim 10^{14}\,$kg if they come within $d \lesssim 1\,$AU of either TM. The focusing effect enhances the number of trajectories passing within $1/m_\phi$ of the Sun (see \figref{fig:perihelionPlot}, where we used $1/m_\phi = 3\,$AU). 
However, \Eros~and \Thomsen~have orbits with average distances of 1.5\,AU and 2.3\,AU from the Sun, respectively; this TM configuration thus sees a much lower rate of clumps passing within $1\,$AU of the detector than the ``Sun-straddling 1\,AU'' TM configuration where both TMs are at fixed positions $0.5\,$AU from the Sun. 
On the other hand, for $\mcl \gtrsim 10^{14}\,$kg, our projections for Eros--Thomsen are slightly more optimistic than for the ``Sun-straddling 1\,AU'' case: in this regime, the detectors are sensitive to clumps passing with $d \gtrsim 1\,$AU, making the focusing effect less relevant; instead, the Eros--Thomsen case benefits from its longer baseline ($\bar{L} \sim 3.1$\,AU) as discussed in Sec.~\ref{sect:MF}. 
We have checked explicitly that the suppression of the rate $\dot{N}_{\cl}$ for $\mcl \lesssim 10^{14}\,$kg for the Eros--Thomsen case is not due to an unfortunate choice of TMs; different choices for the asteroids used as TMs (e.g., replacing \Thomsen\ with 1627~Ivar; see discussion in \citeR{Fedderke:2021kuy}) lead to similar results. 

The effect of the clump-trajectory focusing from the fifth force between the clumps and the Sun suggest that TMs such as Eros--Thomsen which spend most of their orbits at distances of a few AU from the Sun would be more sensitive to the case where the range of the force is larger, such that the clump trajectories are focused into a larger region in the inner Solar System. The $\dot{N}_{\cl}(\mcl)$ projections for the Eros-Thomsen TM configuration and fifth-force ranges of $1/m_\phi = \{1, 3, 5\}\,$AU (keeping $\xi = 10^{3}$ fixed) indeed reflect this expected behavior.
The larger the range of the fifth force, the more clump trajectories get focused into a region where the TMs spend most of their time on their orbits.
It is expected that the rate of this enhancement in $\dot{N}_{\cl}(\mcl)$ with increasing $1/m_\phi$ slows down as the range of the force becomes larger than the typical distance of the TMs orbits from the Sun; our results reflect this trend.

To demonstrate the effect of different fifth-force strengths on our $\dot{N}_{\cl}(\mcl)$ projections, we also show results for $\xi = 10^{2}$ and $1/m_\phi = 1\,$AU.
Comparing to the case with $\xi = 10^{3}$ and $1/m_\phi = 1\,$AU, we find that $\dot{N}_{\cl}$ is significantly reduced for smaller $\xi$.
The weaker the fifth force, the closer clumps of a given $\mcl$ have to come to the TMs to give rise to a discoverable acceleration signal; in turn, the number of trajectories coming close enough to the TMs to give rise to discoverable signals becomes smaller. 
This effect is reduced at large clump masses where the detector is sensitive to clumps passing with $d \gg 1/m_\phi$; i.e., where the clump--TM fifth force is exponentially suppressed by the mass of the mediator.
In this regime, our results become independent of $\xi$, because at larger and larger $d \gg 1/m_\phi$, more and more of the detectable trajectories never pass through the region where the fifth-force is appreciably large.
Note that while the weaker fifth force also changes the focusing effect from the Sun--clump interaction, the corresponding re-distribution of the trajectories has a reduced but non-negligible effect on the Eros-Thomsen sensitivity for the value of $1/m_\phi = 1\,$AU we have chosen here: this change mainly affects orbits much closer to the Sun than the orbits of \Eros~and \Thomsen.
However, because of the modification to the trajectory-focusing effect that nevertheless accompanies a change in $\xi$, the change in $\dot{N}_{\cl}$ at $\mcl \lesssim 10^{14}\,$kg when comparing the results for $\xi = 10^2$ and $\xi = 10^3$ is larger than one in direct proportion to the change in $\xi$: changing the density of the clumps to keep $\kappa \xi$ constant would thus not lead to a constant $\dot{N}_\cl$ prediction as one might have na\"{i}vely expected from \eqref{eq:Rate} (with the necessary modifications to the fifth-force case).

In this section, we have demonstrated that an additional fifth force between clumps and ordinary matter would give rise to multiple non-trivial effects, leading to interesting implications on the sensitivity of GW detectors to the passage of the clumps.
Trivially, additional interactions make it easier to detect such clumps: we found expected rates of discoverable signals with a fifth force present to be much larger than for the gravity-only case, despite the much smaller clump density that we assumed for our fifth-force results. 
Perhaps less obviously, the new force also leads to focusing effects on the clump trajectories which, for the non-trivial velocity distribution of the clumps we consider here, can only be captured by a numerical simulation of trajectories of the type we have undertaken. 
Furthermore, the interplay of the strength of the fifth force, its range, and the location of the TMs leads to relevant and non-trivial changes to the expected rate of discoverable signals, as exemplified by the different benchmark cases for these parameters we studied here.

\section{Conclusions}
\label{sect:conclusions}

A plethora of GW detectors with coverage over an enormous range of frequencies from nHz all the way up to MHz are already operating, under development, or proposed. 
The GW science cases for these detectors is well-established, compelling, and further augmented by the astrophysical and cosmological mysteries upon which they may shed light.
That these detectors can also be sensitive to new physics in the dark sector is still further motivation to push forward with this observational program.

In this work, we studied the sensitivity of GW detectors with test masses located in the inner Solar System to the passage of dark clumps.
One particular realization of such clumps may be DM.
Given our present (lack of) knowledge of the nature of DM, it is entirely plausible that (some fraction) is in the form of such exotic objects.

We described an heuristic estimate of the sensitivity of GW detectors operating in different frequency ranges to the passage of clumps in different mass ranges.
The results of this study are summarized in \figref{fig:OoM_results}.
Given that the inner Solar System is of $\sim \text{AU}$ size and DM clump travel at speeds $v_{\cl} \sim 10^{-3}c$, leading to transit times on the order of $\sim 10^6$\,s, it is perhaps unsurprising that we conclude that GW detectors with sensitivity around the $\mu$Hz band are the most promising for this search.
Such detectors are most sensitive to clumps in the asteroid-mass range around $m_{\cl} \sim 10^{14}$\,kg.
This heuristic estimate, however, intentionally simplified away many details of the clump transits, such as their distribution of orientations relative to the GW detector and their non-trivial velocity distribution; our results here would thus be open to question unless we went further.

We thus also performed a more sophisticated estimation of the sensitivity of a specific GW-detection proposal whose sensitivity is peaked in the $\mu$Hz band, the asteroid-to-asteroid ranging idea advanced in \citeR{Fedderke:2021kuy}, to dark clumps.
This more sophisticated calculation accounts for the distribution of clump trajectories.
We evolved randomly selected clump trajectories through the inner Solar System under the influence of forces sourced by the Sun and computed the signal-to-noise ratio of the acceleration signal induced in the GW detector using a matched-filter approach. 

We first performed this computation for purely gravitational interactions of the clumps with the detector TMs and the Sun, finding good agreement with the results of our earlier heuristic computation, but also demonstrating that the heuristic computation was too conservative in some ways.
The results of this analysis are shown in \figref{fig:gravPlot}.
Nevertheless, the most optimistic rate projections for this refined computation, assuming the clumps comprise all of the average local DM density, indicate that about 1 event in 20 years would be borderline detectable in such a GW detector if the clump mass was $\mcl \sim 10^{14}$\,kg.
It is possible that the DM density in the close vicinity of the Solar System could be enhanced by, e.g., DM substructure, which would increase the event rate.
Additionally, should future-generation GW detector sensitivities be improved, larger event rates and sensitivity to wider DM mass ranges via purely gravitational interactions may be enabled.

We also considered the modifications to this picture that would be induced were there to exist a new, attractive, long-range, stronger-than-gravity fifth force between the dark sector and ordinary matter.
Such a force can be constructed to easily evade all known fifth-force and DM self-interaction constraints, and it leads to dramatic modifications to the detectability prospects for clumpy DM using local-TM-based GW detectors.
The fifth force influences the detectability of clumps both directly by increasing the clump--TM interaction strength as compared to gravity, and indirectly as a result of the strengthened clump--Sun interactions that result in strong focusing effects on clumps transiting the inner Solar System (see \figref{fig:perihelionPlot}).
Within the context of the asteroid-to-asteroid ranging proposal of \citeR{Fedderke:2021kuy}, and making use of realistic asteroid TMs such as \Eros\ and \Thomsen\, we find that rate of detectable clump transits around $10^{10}\,\text{kg}\lesssim \mcl \lesssim 10^{14}\,\text{kg}$ could be as high as a few per year even if the clumps are only a $\sim 1\,\%$ sub-component of the local average DM density; see \figref{fig:FFPlot}.
These results hold for a clump--SM fifth-force with a range of $1$--$(\text{few})$\,AU and $10^3$ times stronger than gravity.
We also studied the parametric dependence of these results to changing fifth-force strength and range.

We conclude that local-TM-based $\mu$Hz GW detectors in the inner Solar System hold promise to probe the local passage of heavy (composite) dark states imbued with stronger-than-gravity fifth-force couplings to the SM, over a fairly wide range of parameters.

\acknowledgments
We thank Surjeet Rajendran for fruitful discussions.
We are grateful to Harikrishnan Ramani for pointing out a typographical error in an earlier version of the paper.
This work was supported by the Simons Investigator Award No.~824870, NSF Grant No.~PHY-2014215, DOE HEP
QuantISED Award No.~100495, the Gordon and Betty Moore Foundation Grant No.~GBMF7946, and the U.S.~Department of Energy (DOE), Office of Science, National Quantum Information Science Research Centers, Superconducting Quantum Materials and Systems Center (SQMS) under contract No.~DE-AC02-07CH11359. 
The work of M.A.F.~was performed in part at the Aspen Center for Physics, which is supported by National Science Foundation (NSF) Grant No.~PHY-1607611. 


\appendix

\section{The Improved Heuristic Estimate}
\label{app:matchedFilterScaling}
We sketch here the origin of the replacement $h_c(f) \rightarrow (\tilde{f}/f)\times h_c(\tilde{f})$ for \eqref{eq:mcmin} that led to the improved heuristic estimate given in \sectref{sect:OoM}.

Consider that the discrete, finite-time analogue of the matched filter (squared) SNR at \eqref{eq:SNR} is given by
\begin{align}
    \rho^2 \sim \sum_k \frac{4}{T} \frac{|\tilde{a}_k|^2}{S_a(f_k)}\:,
    \label{eq:SNRapp}
\end{align}
where we have omitted the zero-frequency term and possible Nyquist term, taken an arbitrary duration-$T$ interval, and set $f_k = k/T$; see also \citeR{Fedderke:2020yfy}.

We wish to relate $\tilde{a}_k$ to the known size of the time-domain acceleration signal.

We take as input to this discussion the following definitions and relationships taken from, e.g., Appendix C of \citeR{Fedderke:2020yfy}, whose conventions we have adopted throughout this work: 
\begin{enumerate}
    \item\label{pt:PSDdefn} we can \emph{define}%
    \footnote{\label{ftnt:ZeroNyquist}%
        The zero- and Nyquist-frequency terms have different definitions, but those are not important to this discussion; see \citeR{Fedderke:2020yfy}.
        } %
    the one-sided PSD of any discretely sampled (real) time-domain signal $a(t_n)$ whose discrete Fourier transform data are $\tilde{a}(f_k)$ as $\mathcal{S}(f_k) \equiv (2/T) |\tilde{a}(f_k)|^2$; and 
    \item\label{pt:PSDtime} the one-sided PSD is related to the time-average of the time-domain signal via $\langle |a(t)|^2 \rangle_T = T^{-1} \sum_k \mathcal{S}_a(f_k)$; together with point \ref{pt:PSDdefn}, this is just a statement of Parseval's theorem.
\end{enumerate}

As another important input, recall that we observed in \sectref{sect:OoM} that the PSD for the acceleration signal induced by a clump passage is roughly flat for $f \lesssim f_{\text{peak}}$: $\mathcal{S}_a(f\lesssim f_{\text{peak}}) \sim \mathcal{S}_a^0$.
We also noted that it falls off exponentially for $f \gtrsim f_{\text{peak}}$~\cite{LISA-Pre-Phase-A}.

Substituting this into the relationship at point \ref{pt:PSDtime}, and noting that the DFT frequency spacing is $\Delta f \equiv 1/T$, we can write%
\footnote{\label{ftnt:meanSubtractedTimeSeries}%
    Technically, we assume here that $a(t)$ is the mean-subtracted acceleration time series, but this changes the argument by only $\mathcal{O}(1)$ factors.
    } %
$\langle |a(t)|^2 \rangle_T \sim \mathcal{S}_a^0 \sum_{k | f_k<f_{\text{peak}}} \Delta f \sim f_{\text{peak}} \mathcal{S}_a^0$.
But we also separately know the magnitude of $a(t)$: approximately, we have that $|a(t)| \sim a_0 \sim G_N ( m_{\cl} / d^2 ) \times \min\lb[ 1 , 2 L/d \rb]$ for a duration of time $\Delta t \sim 1/f_{\text{peak}}$ during the arbitrary length-$T$ interval over which the time-series data are assumed to be available for analysis.
We can thus estimate $\langle |a(t)|^2 \rangle_T \sim (\Delta t / T) \times a_0^2 \sim  a_0^2 / (f_{\text{peak}}T)$.
Combining this with the previous result, we have $\mathcal{S}_a^0 \sim a_0^2 / (f_{\text{peak}}^2T)$.
Then, invoking the definition at point \ref{pt:PSDdefn}, we arrive at the desired relationship between the Fourier amplitude of the signal and the time-domain amplitude: $|\tilde{a}(f_k)|^2 \sim a_0^2 / (2f_{\text{peak}}^2)$ for $f_k \lesssim f_{\text{peak}}$, and $|\tilde{a}(f_k)|^2 \sim 0 $ for $f_k \gtrsim f_{\text{peak}}$.

Also, recall from \eqref{eq:SaRelationship} that $f_k S_a(f_k) \sim  (2\pi f_k)^4 L^2 [h_c(f_k)]^2/2$.

Putting this all into \eqref{eq:SNRapp}, we obtain
\begin{align}\label{eq:rho2}
    \rho^2 \sim \sum_{k|f_k\lesssim f_{\text{peak}}} 4 \Delta f  \frac{a_0^2}{f_{\text{peak}}^2 (2\pi)^4 f_k^3 L^2 [h_c(f_k)]^2}\:,
\end{align}
where have again used $\Delta f = T^{-1}$ to rewrite the factor of $T^{-1}$ appearing explicitly in \eqref{eq:SNRapp}.

Now, suppose $h_c(f_k)$ is monotonically decreasing on $f_k \lesssim f_{\text{peak}}$, and assume that $h_c(f_k) \sim h_c(f_{\text{peak}})$ over a bandwidth $\delta f \sim f_{\text{peak}}$, which is typical.
Then we would have the summand in \eqref{eq:rho2} roughly constant over that bandwidth and negligible otherwise, along with $f_k \sim f_{\text{peak}}$ and $\sum \Delta f \sim f_{\text{peak}}$, leading to 
\begin{align}
    \rho \propto \frac{a_0}{f_{\text{peak}}^2 L h_c(f_{\text{peak}})}\:,
\end{align}
where we have now dropped numerical constants and focus on parametric dependences only. 
Fixing $\rho \sim 1$ then yields the minimum detectable acceleration amplitude $a^{\text{min}}_0 \propto f_{\text{peak}}^2 L h_c(f_{\text{peak}})$, leading to 
\begin{align}
    m_{\cl}^{\text{min}} \propto \frac{v_{\cl}^2 h_c(f_{\text{peak}}) L}{G_N} \times \max\lb[ 1 , \frac{4\pi f_{\text{peak}} L}{v_{\cl}} \rb]\:,
\end{align}
which has exactly the same parametric dependence as \eqref{eq:mcmin}.
The $\mathcal{O}(1)$ factors can be similarly recovered.

However, if $h_c(f_k)$ has a broad minimum such that $fh_c(f)$ is minimized for $f\sim \tilde{f} \lesssim f_{\text{peak}}$, with $fh_c(f) \sim \tilde{f}h_c(\tilde{f})$ over a typical bandwidth $\delta f \sim \tilde{f}$, then the parametrics change, and instead we have
\begin{align}
    \rho \propto \frac{a_0}{f_{\text{peak}} \tilde{f} L h_c(\tilde{f})}\:,
\end{align}
again dropping numerical factors and keeping only parametric dependences.
Fixing $\rho \sim 1$ now gives the result $a^{\text{min}}_0 \propto f_{\text{peak}}\tilde{f} L h_c(\tilde{f})$, leading now to 
\begin{align}
m_{\cl}^{\text{min}} \propto \frac{v_{\cl}^2 \lb[ \tilde{f} h_c(\tilde{f}) / f_{\text{peak}} \rb] L}{G_N} \times \max\lb[ 1 , \frac{4\pi f_{\text{peak}} L}{v_{\cl}} \rb]\:,
\end{align}
which, up to $\mathcal{O}(1)$ factors we have not tracked here, gives the replacement rule we stated for \eqref{eq:mcmin} given that we wrote $f\sim f_{\text{peak}}$ implicitly there.

\section{Some Technical Simulation Details}
\label{app:technicalDetail}

In this appendix, we outline some more technical details of our Monte Carlo and trajectory computation simulation discussed in \sectref[s]{sect:MF} and \ref{sect:FF}.

\subsection{A `Trick' to Avoid Spectral Leakage}
\label{app:spectralLeakageTrick}
In \sectref{sect:MF}, we noted that we utilized a $T=10\,$yr mission duration for our simulation.
However, this presents a subtlety.
A clump passage whose moment of closest approach $t_d$ and duration of transit $T_{\text{peak}} \sim 1/f_{\text{peak}}$ are such that a hard top-hat window of duration $T=10\,$yrs would clip off a part of the acceleration time series $\Delta a_{ij}(t)$ defined at \eqref{eq:BPDA} while it is still non-negligibly small, would give rise to artificial discontinuities and result in spectral leakage effects that would distort our results; see discussion in \citeR{Harris:1978wdg}, and Appendix D of \citeR{Fedderke:2020yfy}.
That is, we would mis-model the acceleration time series if $0 \leq t_d < T_{\text{peak}}$ or $T-T_{\text{peak}}\lesssim t_d \leq T$.

Instead of windowing the data to mitigate this, which would be a more standard approach, we can perform a simple `trick' and impose approximately and by hand at the time-series level the signal periodicity that a discrete Fourier transform (see footnote \ref{ftnt:discreteFT}) assumes of its input signal: namely, we replace
\begin{align}
    \Delta a_{ij}(t) \rightarrow  \Delta a_{ij}(t) + \Delta a_{ij}(t+T) + \Delta a_{ij}(t-T) \:.
    \label{eq:trickAccn}
\end{align}
For a single transit of a clump with $t_d$ outside the ranges mentioned above, this has negligible impact on the results so long as $T_{\text{peak}} \ll T$, as the latter two terms on the RHS of \eqref{eq:trickAccn} are then negligible compared to the first.
For transits with $t_d \sim 0,T$, however, this substitution has the effect of wrapping around the part of the acceleration time series that would otherwise be clipped off so that it appears near $t\sim T,0$, respectively.
While this is of course unphysical at the level of the time series, that is not relevant in our computation of the PSD; the key point is that this procedure results in a computation of the signal PSD that is unaffected by unphysical spectral leakage effects imposed by a hard mission-duration cutoff because the DFT automatically imposes signal periodicity outside its defining domain.

There is a further sampling effect that this trick addresses. 
In our simulation, we sample values of $t_0$, the TI-sphere \emph{entrance} crossing time in the range $t_0 \in [0,T]$, but we also assume that the mission duration is $t\in [0,T]$.
Therefore, for the first duration of time $0\lesssim t \lesssim ( R_0 - r_{\text{det}} ) / v_{\cl} $ where $r_{\text{det}}$ is a typical detector heliocentric distance, there would be an artificial deficit of clumps passing the detector TMs at small $d$.
Likewise, clumps entering the TI sphere during the final amount of time $ T - ( R_0 - r_{\text{det}} ) / v_{\cl} \lesssim t \lesssim T$ do not approach the detectors closely enough to achieve small values of $d$ before the end of the simulation.
For some parameters this could in principle be quite a severe issue: with $R_0 \sim 37\,$AU and $v_{\cl} \sim 10^{-3}c$, $R_0/(v_{\cl}T) \sim 0.6$.
Adding in the latter two terms on the RHS of \eqref{eq:trickAccn} corrects both these effects by (a) effectively extending the simulation end time by an additional amount of time $T$, which is broadly sufficient to capture the full duration of the latter passages through the TI sphere (except for very slow clumps), and (b) wrapping those transits around to populate the equal-length interval at the beginning of the simulation which is artificially depleted by our sampling approach.
While this is not at all relevant for the Sun-straddling, spatially fixed baselines, it is relevant for the Eros--Thomsen and similar baselines we consider, as these evolve in time.
It does however give rise to a sampling of orbital configurations of the TMs in our simulation that is somewhat offset in time from the nominal mission interval $t\in[0,T]$; however, this is degenerate with a mission start-time selection effect, and we have already assumed an arbitrary mission start time.

Note that for transits with $T_{\text{peak}} \gtrsim T$, the above procedures would yield inaccurate results because the implicit assumption in performing this trick is that the acceleration time series goes to zero well within the simulation duration, so that at most one of the terms on the RHS of \eqref{eq:trickAccn} is non-negligible at a fixed $t$.
However, with $T=10\,$yrs chosen, these inaccurate trajectory results are out of the band of sensitivity of the asteroid-ranging detector, and this is therefore of little to no physical relevance to our results.

\subsection{Extraction of Rate from Simulations}
\label{app:rateExtraction}

In this appendix, we detail how we extract our rate estimates from our numerical simulations.

As discussed in \sectref{sect:MF}, we perform a Monte Carlo (MC) sampling of initial conditions for the clump trajectories drawn from appropriate distributions of location, speed, and direction for the clumps, assuming that clump trajectories are initialized on the trajectory-initialization (TI) sphere:%
\footnote{\label{ftnt:sampledVariables}%
    To be precise, for each trajectory we sample the location of entrance of the trajectory on the TI sphere (2 parameters), an inward-directed velocity of the particle on the TI sphere (3 parameters), and a time of TI-sphere entrance $t_0$ (1 parameter).} %
a Sun-centered sphere of radius $R_0$.
The assumed underlying velocity distribution of the clumps is that of the Standard Halo Model~(SHM)~\cite{Evans:2018bqy}, a truncated Maxwell--Boltzmann distribution boosted in the direction of Cygnus, assumed to be located at ecliptic longitude $\sim 331^\circ$ and ecliptic latitude $\sim 57.5^\circ$.
Appropriate flux-weighted sampling on a boosted sphere is non-trivial, but can be done quasi-analytically: sequentially marginalized cumulative distribution functions (CDFs) for all relevant physical parameters can be derived in closed form (these CDF expressions are all algebraically complicated, and we omit them here).
The only MC piece of the computation then involves sequential uniform random sampling of marginalized CDF values on the interval [0,1], followed by numerical inversion of the marginalized CDFs to obtain the requisite physical parameter samples.%
\footnote{\label{ftnt:sequentialMarginalization}%
    For example, we mean the following procedure. 
    Suppose we have a differential distribution $d^2N/ (dx dy)$ of a number of events $N$ that has a known parametric dependence on $x$ and $y$. 
    We can integrate out / marginalize over $y$ by computing $dN/dx \equiv \int dy [ d^2N/ (dx dy)]$. 
    This can be turned into a CDF for $x$: $C[x] \equiv [ \int^x dx' ( dN / dx' ) ] / [ \int dx' ( dN / dx' ) ]$.
    If we generate $r \sim \mathcal{U}[0,1]$ and solve $r = C[x]$, we obtain a sample $x = C^{-1}[r]$.
    Then we can return to the differential distribution $d^2N/ (dx dy)$, condition on $x =  C^{-1}[r]$, and define a CDF for $y$ conditional on $x= C^{-1}[r]$: $D[y|x= C^{-1}[r]] = \int^y dy' [ d^2N/ (dx dy)|_{x= C^{-1}[r]}] / \int dy' [ d^2N/ (dx dy)|_{x= C^{-1}[r]}]$, and extract a sample for $y$ by generating $s \sim \mathcal{U}[0,1]$ and solving $D[y|x= C^{-1}[r]] = s$ for $y = D^{-1}[s;x= C^{-1}[r]]$. This procedure can be repeated sequentially for more than two parameters, with each random sample from $\mathcal{U}[0,1]$ involving `peeling back' one more `layer' of the differential distribution.} %
For each set of parameters at the TI sphere, we numerically solve for the corresponding orbits under the action of gravity (and, if applicable, the fifth force), taking into account only Sun--TM interactions.
We assume the Sun is a spatially fixed object, and ignore the effects of other bodies in the Solar System.
For each trajectory $k$, we then perform the baseline-projected acceleration computation discussed in \sectref{sect:MF}, and solve for the minimum mass $m_{\cl,k}^{\text{min}}$ required on the computed trajectory to obtain an SNR $\rho_k = 1$ in the matched-filter search.

In performing our computations for the Eros--Thomsen TM baseline assumption, we assume the TMs follow the elliptical orbits specified by the osculating orbital elements of \Eros\ and \Thomsen\ as given in the JPL Small-Body Database~\cite{JPL-SBD}; see \tabref{tab:ErosThomsenOrbitalElements}.
We follow the procedures of Appendix A.4 of \citeR{Fedderke:2020yfy} to generate the locations of the TMs as a function of time.%
\footnote{\label{ftnt:typoInOtherPaper}%
    We note a minor typographical error in the text just below Eq.~(A18) in \citeR{Fedderke:2020yfy}: `clockwise rotation' should read instead `counter-clockwise rotation'. 
    That is, the sense of rotation is such that the indicated (active) $SO(3)$ rotation matrix around, e.g., the $z$ axis by an angle $+\theta$ should have `$-\sin\theta$' appearing in the row 1, column 2 position, assuming that rotation is implemented by matrix multiplication on the left of a column vector representing the co-ordinates of an object in a fixed frame.} %

\begin{table}[t]
    \begin{ruledtabular}
        \caption{\label{tab:ErosThomsenOrbitalElements}%
            Osculating elliptical orbital elements for \Eros\ and \Thomsen, as retrieved from the JPL Small-Body Database (SBD)~\cite{JPL-SBD} on March 8, 2022, and utilized to generate our ``Eros--Thomsen'' TM trajectories.
            We give the semi-major axis $a$, eccentricity $e$, orbital inclination from the ecliptic $i$, longitude of the ascending node $\Omega$, and argument of perihelion $\omega$.
            All angles are expressed in degrees.
            For \Eros, we treat one selected time of perihelion passage $t_p$ as an arbitrary parameter centered on our assumed duration-$T$ mission simulation; we however treat $t_p$ for \Thomsen\ with the correct relative offset as compared to that of \Eros.
            These parameters are shown here in rounded form; most are known to (much) higher precision. In our computation, we utilize these parameters with their full known precision, although we note that, for these objects, over timescales much longer than our simulation, the osculating orbit approximation breaks down to a relevant degree. 
            }
    \begin{tabular}{lll}
    Element         &   \Eros       &   \Thomsen    \\ \hline
    $a$ [AU]        &    1.45827    &   2.17884     \\
    $e$             &    0.22273    &   0.32972     \\
    $i$ [degrees]       &    10.8285    &   5.69482     \\
    $\Omega$ [degrees]  &    304.296    &   302.136     \\
    $\omega$ [degrees]  &    178.897    &   2.78751     \\
    $t_p$ [s]       &   \text{arbitrary; see caption} &   $t_p^{\text{Eros}} + 7.16 \times 10^6 $    
    \end{tabular}
    \end{ruledtabular}
\end{table}

For computational efficiency, we split our MC simulation into sub-simulations. Enumerating the sub-simulations as $i = 1, \ldots, i_{\rm max}$, we consider disjoint ranges of the closest TM-clump approach distance, $d$, as  $d \in ( d_i^{\text{min}}, d_i^{\text{max}} ] \equiv \mathcal{D}_i$. The boundaries of the $\mathcal{D}_i$ regions we choose are as indicated by the vertical dotted black lines in the left panels of \figref[s]{fig:gravPlot} and \ref{fig:FFPlot}, except $d_1^{\rm min} = 1.5 \times 10^{-3}\,$AU and $d_{i_{\rm max}}^{\rm max} = R_0$. Note that the $d_1^{\rm min}$ cut is imposed for computational reasons; simulating trajectories with smaller $d$ requires exceedingly resource-intensive computations, and our results verify {\it a posteriori} that the chosen value, $d_1^{\rm min} = 1.5 \times 10^{-3}\,$AU, is sufficiently small to not affect our numerical results. 

For each sub-simulation, we draw $N_i^{\rm MC}$ initial conditions at the TI sphere, and calculate the corresponding orbits in the gravitational (and, if applicable, fifth-force) potential of the Sun; note that typically we use larger and larger $N_i^{\rm MC}$ for smaller and smaller $d_i^{\rm max}$ to obtain sufficient statistics. From the calculated orbits, we then select the orbits with $d \in \mathcal{D}_i$. For each such selected orbit, we perform the baseline-projected acceleration computation discussed in \sectref{sect:MF} and calculate the SNR assuming a reference value of $\mcl = \mcl^{\rm ref}$; since the acceleration signal scales linearly with $\mcl$, we can trivially obtain the SNR of the $k$-th trajectory for any other clump mass as $\rho_k(\mcl) = \left( \mcl / \mcl^{\rm ref} \right) \times \rho_k(\mcl^{\rm ref})$. Armed with these results, we compute the number of simulated trajectories in the $i$-th sub-simulation that give rise to a detectable signal, defined as $\rho_k(\mcl) \geq 1$ throughout this work, for a given clump mass, $N_{i, {\rm  detectable}}^{\rm MC}(\mcl)$. 

In order to compute the rate of discoverable signals, we need the physical flux-weighted rate, $\dot{N}_{\rm cl}^{\rm physical}$ of clumps expected to (on average) enter the TI sphere for a given clump density $\rho_{\rm cl}$ and velocity distribution of the clumps at the TI sphere. For the homogeneous $\rho_{\rm cl}$ and the SHM velocity distribution we assume in this work, $\dot{N}_{\rm cl}^{\rm physical}$ is straightforward (if laborious) to compute in closed form; the final expressions are algebraically complicated and we omit them here. The (average) physical rate of discoverable signals for a given clump rate is then obtained by summing over the results of the sub-simulations,
\begin{align*}
    \dot{N}_{\cl, {\rm detectable}}^{\rm physical}(\mcl) = \sum_i \left[ \frac{N_{i, {\rm detectable}}^{\rm MC}(\mcl)}{N_i^{\rm MC}} \right] \times \dot{N}_{\rm cl}^{\rm physical} \:.
\end{align*}
The uncertainty of this calculation is obtained from the Poisson error of $N_{i, {\rm detectable}}^{\rm MC}(\mcl)$,
\begin{align*}
    \sigma\left[ N_{i, {\rm detectable}}^{\rm MC}(\mcl) \right] = \sqrt{N_{i, {\rm detectable}}^{\rm MC}(\mcl)} \:;
\end{align*}
since our sub-simulations are statistically independent and disjoint, standard error propagation yields
\begin{align*}
    \frac{ \sigma\left[\dot{N}_{\cl, {\rm detectable}}^{\rm physical}(\mcl)\right]}{\dot{N}_{\rm cl}^{\rm physical}} = \sqrt{ \sum_i \left[ \frac{\sqrt{N_{i, {\rm detectable}}^{\rm MC}(\mcl)}}{N_i^{\rm MC}} \right]^2 }\:.
\end{align*}
\vspace{0.01cm}

\bibliographystyle{JHEP.bst}
\bibliography{theBib}

\end{document}